\documentclass[aps, prb, twocolumn, superscriptaddress, longbibliography]{revtex4-1}
\usepackage{graphicx}
\usepackage{bm}
\usepackage{natbib}
\usepackage{upgreek}
\usepackage{amsmath}
\usepackage{amssymb}

\usepackage{color}
\usepackage[colorlinks,bookmarks=false,citecolor=darkblue,linkcolor=blue,urlcolor=blue]{hyperref}
\definecolor{darkblue}{rgb}{0,0.02,0.45}

\newcommand*{\Mn}{Mn$_2$Mo$_3$O$_8$}
\newcommand*{\vect}[1]{\mathbf{#1}}

\begin{document} 

\title{Magnetic anisotropy and exchange paths for octa- and tetrahedrally coordinated Mn$^{2+}$ ions in the honeycomb multiferroic Mn$_2$Mo$_3$O$_8$}

\author{D. Szaller}
\affiliation{Institute of Solid State Physics, Vienna University of
Technology, 1040 Vienna, Austria}
\author{K. Sz\'asz}
\affiliation{Department of Physics, Budapest University of Technology and Economics, 1111 Budapest, Hungary}
\author{S. Bord\'acs}
\affiliation{Department of Physics, Budapest University of Technology and Economics, 1111 Budapest, Hungary}
\affiliation{Hungarian Academy of Sciences, Premium Postdoctor Program, 1051 Budapest, Hungary}
\author{J. Viirok}
\author{T. R{\~o}{\~o}m}
\author{U. Nagel}
\affiliation{National Institute of Chemical Physics and Biophysics,
Akadeemia tee 23, 12618 Tallinn, Estonia}
\author{A. Shuvaev}
\author{L. Weymann}
\author{A. Pimenov}
\affiliation{Institute of Solid State Physics, Vienna University of
Technology, 1040 Vienna, Austria}
\author{A. A. Tsirlin}
\author{A. Jesche}
\affiliation{Experimental Physics VI, Center for Electronic Correlations and Magnetism, University of Augsburg, 86159 Augsburg, Germany}
\author{L. Prodan}
\affiliation{Institute of Applied Physics, MD-2028 Chi\c{s}in\u{a}u, Republic of Moldova}
\affiliation{Experimental Physics V, Center for Electronic Correlations and Magnetism, University of Augsburg, 86159 Augsburg, Germany}
\author{V. Tsurkan}
\affiliation{Institute of Applied Physics, MD-2028 Chi\c{s}in\u{a}u, Republic of Moldova}
\affiliation{Experimental Physics V, Center for Electronic Correlations and Magnetism, University of Augsburg, 86159 Augsburg, Germany}
\author{I. K\'ezsm\'arki}
\affiliation{Experimental Physics V, Center for Electronic Correlations and Magnetism, University of Augsburg, 86159 Augsburg, Germany}

\begin{abstract}
We investigated the static and dynamic magnetic properties of the polar ferrimagnet \Mn~in three magnetically ordered phases via magnetization, magnetic torque, and THz absorption spectroscopy measurements. The observed magnetic field dependence of the spin-wave resonances,  including Brillouin zone-center and zone-boundary excitations, magnetization, and torque, are well described by an extended two-sublattice antiferromagnetic classical mean-field model. In this orbitally quenched system, the competing weak easy-plane and easy-axis single-ion anisotropies of the two crystallographic sites are determined from the model and assigned to the tetra- and octahedral sites, respectively, by \emph{ab initio} calculations. 
\end{abstract}

\date{\today}


\maketitle

\section{Introduction}

Static magneto-electric (ME) coupling, namely the potential to electrically manipulate magnetic states and magnetically control electric polarization, has opened a new path for data storage\cite{kimura_nature_2003,fiebig_jpd_2005,spaldin_science_2005,eerenstein_nature_2006,cheong_nmat_2007,wu_prl_2013,dong_advph_2015,fiebig_natrev_2016,kuzmenko_prl_2018,weymann_npjqm_2020}. At finite frequencies, the same cross-coupling leads to fascinating optical phenomena\cite{szaller_psr_2019}, such as one-way transparency\cite{kezsmarki_prl_2011, bordacs_nphys_2012,takahashi_nphys_2012,takahashi_prl_2013,szaller_prb_2013,kezsmarki_nc_2014,szaller_prb_2014,kuzmenko_prb_2015,kezsmarki_prl_2015},  reciprocal\cite{bordacs_nphys_2012} and non-reciprocal\cite{kuzmenko_prb_2014,kurumaji_prl_2017} optical rotation. Since these ME phenomena only emerge in systems simultaneously lacking the time-reversal and spatial inversion symmetries, they have been realized in magnetically ordered phases with broken inversion symmetry\cite{kezsmarki_prl_2011, bordacs_nphys_2012,takahashi_nphys_2012,takahashi_prl_2013,kezsmarki_nc_2014,szaller_prb_2014,kuzmenko_prb_2014,kuzmenko_prb_2015,kezsmarki_prl_2015,kurumaji_prl_2017}, and in the paramagnetic phase of non-centrosymmetric compounds when a magnetic field was applied\cite{yu_prl_2018,viirok_prb_2019,kuzmenko_prb_2019,yokosuk_npjqm_2020}. 

In most compounds\cite{katsura_prl_2005,jia_prb_2007,arima_jpcm_2008,murakawa_prl_2010}, the ME coupling arises from the spin-orbit interaction, 
thus, the strength of the ME coupling is strongly limited by its relativistic origin.  However, in magnetically ordered non-centrosymmetric crystals, the symmetric exchange-striction\cite{sergienko_prl_2006} provides an alternative mechanism to generate ME coupling, which exists even for spin-only ions with half-filled  $d$-shell. Depending on the relative orientation of magnetic moments on crystallographic sites connected by an exchange path, the magnetic order can further be stabilized by distorting the bond and by that modifying the strength of the exchange coupling. This distortion, driven by the magnetic order, also produces electric polarization in non-centrosymmetric crystals, realizing the ME coupling. To create a ME monodomain state with magnetically induced macroscopic polarization, either an electric field is applied\cite{choi_prl_2008}, or pyroelectric polarization is necessary, when cooling the system below the magnetic ordering temperature\cite{wang_scirep_2015}. The latter condition is literally fulfilled in type-I multiferroics\cite{khomskii_jmmm_2006}, where the onset of magnetic order takes place within a pre-existing polar state. 

\begin{figure}[tbp]
\begin{center}
\includegraphics[width=0.5\linewidth, clip]{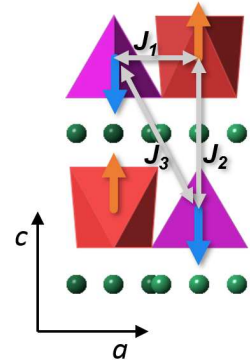}\vspace{-0.8cm}
\end{center}
\caption{\textbf{Crystal and magnetic structure of \Mn.} Red octahedra and purple tetrahedra show the oxygen coordination of Mn atoms, while green spheres represent the Mo atoms. The zero-field ferrimagnetic spin configuration is indicated by orange and blue vectors, and relevant antiferromagnetic exchange paths are shown by grey arrows.} \label{fig_struct}
\end{figure}

The members of the polar hexagonal (space group $P6_3mc$) \emph{M}$_2$Mo$_3$O$_8$ crystal family, where $\emph{M}$ stands for transition metal ions, are ideal candidates for strong, exchange-striction based ME effects\cite{wang_scirep_2015}. The honeycomb $ab$-plane layers of \emph{M} magnetic moments are separated by the Mo$^{4+}$ layers (see Fig.~\ref{fig_struct}), which are non-magnetic in these compounds due to the formation of Mo$_3$O$_{13}$ trimer singlets\cite{cotton_qrcs_1966}. Half of the \emph{M} ions are in octahedral and half in tetrahedral oxygen environment, as presented in Fig.~\ref{fig_struct}. Due to a delicate balance of competing superexchange paths the magnetic ordering of octa- and tetrahedrally coordinated magnetic moments in different layers can lead to various types of spin structures in this material class, such as collinear easy-axis antiferromagnetic\cite{varret_jp_1972}, ferrimagnetic\cite{mcalister_jmmm_1983} and spin-flopped plane\cite{wang_scirep_2015} ordered states. These states can also be transformed into each other by an external magnetic field\cite{wang_scirep_2015,kurumaji_prx_2015,kurumaji_prb_2017}.

At low temperatures, Fe$_2$Mo$_3$O$_8$ presents the largest magnetically switchable electric polarization among single-phase multiferroic crystals\cite{wang_scirep_2015,kurumaji_prx_2015}. The ME susceptibility can be tuned by diluting Fe by Mn\cite{kurumaji_prb_2017} or non-magnetic Zn\cite{kurumaji_prx_2015}. The spin excitations of these compounds, that are classified as magnons and electromagnons\cite{kurumaji_prbr_2017}, show one-way transparency in the paramagnetic phase\cite{yu_prl_2018} and non-reciprocal optical rotation\cite{kurumaji_prl_2017}. However, the microscopic description of the sequence of magnetic phases and the spin-wave resonances in this material family is still an open problem. 

\Mn~offers the perfect starting point to understand the magnetic properties of the \emph{M}$_2$Mo$_3$O$_8$ compounds. The half-filled 3$d$ shells of Mn$^{2+}$ ions with $S=5/2$ spin and $L=0$ orbital moment allow magnetic single-ion anisotropies only via higher-order interactions\cite{watanabe_ptp_1957}. Thus, the resulting magnetic single-ion anisotropies of the tetra- and octahedral sites are expected to be weak. \Mn~has an easy-axis type ferrimagnetic ground state below $T_N=41\textrm{ K}$ where the spins of the octa- and tetrahedral sites are aligned antiparallel along the hexagonal axis of the crystal\cite{mcalister_jmmm_1983}, as shown in Fig.~\ref{fig_struct}. Although the magnetic moments of the two crystallographic sites compensate each other when approaching the lowest temperature, the spontaneous magnetization is finite at higher temperatures indicating the different temperature dependences of the ordered moments at the two sites\cite{mcalister_jmmm_1983,mcalister_jap_1984}. At low temperatures, the magnetization remains zero in magnetic fields  along the hexagonal axis up to $\mu_0 H_{C1}=4\textrm{ T}$, above which it starts to smoothly increase\cite{kurumaji_prb_2017}, showing an evident sign of a spin-reorientation transition.

In order to gain a deeper insight into the microscopic mechanisms governing the magnetic behaviour of \Mn, we followed the magnetic field dependence of spin-wave resonances through three magnetic phases.
The field dependence of the resonance frequencies and the bulk magnetization were successfully described by a simple microscopic model, which can serve as a starting point to understand other systems of the \emph{M}$_2$Mo$_3$O$_8$ crystal family. The key results of the mean-field analysis are also supported by our first principle calculations.

\section{Methods}

\Mn~single crystals were grown by the chemical transport reaction method using anhydrous TeC$_{l4}$ as a transport agent. Plate-like crystals with the dimension of about 2-3\,mm in the $ab$ plane and 0.5-1\,mm along the $c$ axis were obtained after one month transport at 1000\,$^\circ\textrm{C}$ with a temperature difference of 50\,$^\circ\textrm{C}$. 
The magnetization measurements were performed using a Squid magnetometer (MPMS-5, Quantum Design) in fields up to $\mu_0 H=5\textrm{ T}$ and vibrating sample magnetometer in fields up to $\mu_0 H=14\textrm{ T}$ using Physical Properties Measurement System (PPMS, Quantum Design). The torque magnetometry was also performed in a PPMS with magnetic fields of up to 9\,T.
 
Optical transmission experiments between 60 and 180\,GHz were carried out using quasi-optical terahertz spectroscopy~\cite{volkov_infrared_1985}. This technique utilizes linearly polarized monochromatic radiation provided by backward-wave oscillators. Liquid He-cooled bolometer was used as detector of the transmitted radiation.  The sample was in a He-cooled cryostat and the experiments were performed at $T=3\textrm{~K}$ temperature. The magnetic field was parallel to the propagation direction of the light beam (Faraday configuration). In order to increase the sensitivity of the absorption measurement in the frequency range where the sample dimensions $(\sim 1\textrm{ mm})$ are less than the wavelength of light $(\lambda\approx 3\textrm{ mm})$, the experiments were performed in the fixed frequency mode while sweeping the magnetic field in the $\mu_0 H=0-7\textrm{ T}$ range.

Fourier-transform spectroscopy was used to study the optical absorption between 120 and 6000\,GHz with 8\,GHz resolution. The magnetic field dependence of the spectra in magnetic fields up to $\mu_0 H=17\textrm{ T}$  was investigated using the TeslaFIR setup of the National Institute of Chemical Physics and Biophysics in Tallinn.\cite{kezsmarki_nc_2014,fishman_book_2018} This setup consists of a Martin-Puplett interferometer, a mercury arc lamp as a light source and a Si bolometer cooled down to 300\,mK as a light intensity detector. 
The transmission spectra at $T=3\textrm{~K}$ were measured in both the Faraday and Voigt configuration, i.e. in magnetic fields parallel and perpendicular to the direction of light propagation, respectively, using linearly polarized incoming beam and unpolarized detection.

Magnetic anisotropy was estimated by density-functional theory (DFT) band-structure calculations performed in the VASP package~\cite{vasp1,vasp2} for the experimental crystal structure at 1.7\,K~\cite{duncan2020} using generalized gradient approximation for the exchange-correlation potential\cite{perdew_prl_1996}. Correlation effects in the Mn $3d$ shell were included on the mean-field via the DFT+$U$ correction with the Coulomb repulsion parameter $U=5$\,eV and Hund's exchange $\mathcal{J}=1$\,eV~\cite{nath2014}. Single-ion anisotropy for individual Mn sites was calculated from total energies of orthogonal spin configurations as described in Ref.~\onlinecite{xiang2011}. The $g$-factor values were estimated from calculated orbital moments.

\section{Results}

\begin{figure}[tbp]
\begin{center}
\includegraphics[width=0.9\linewidth, clip]{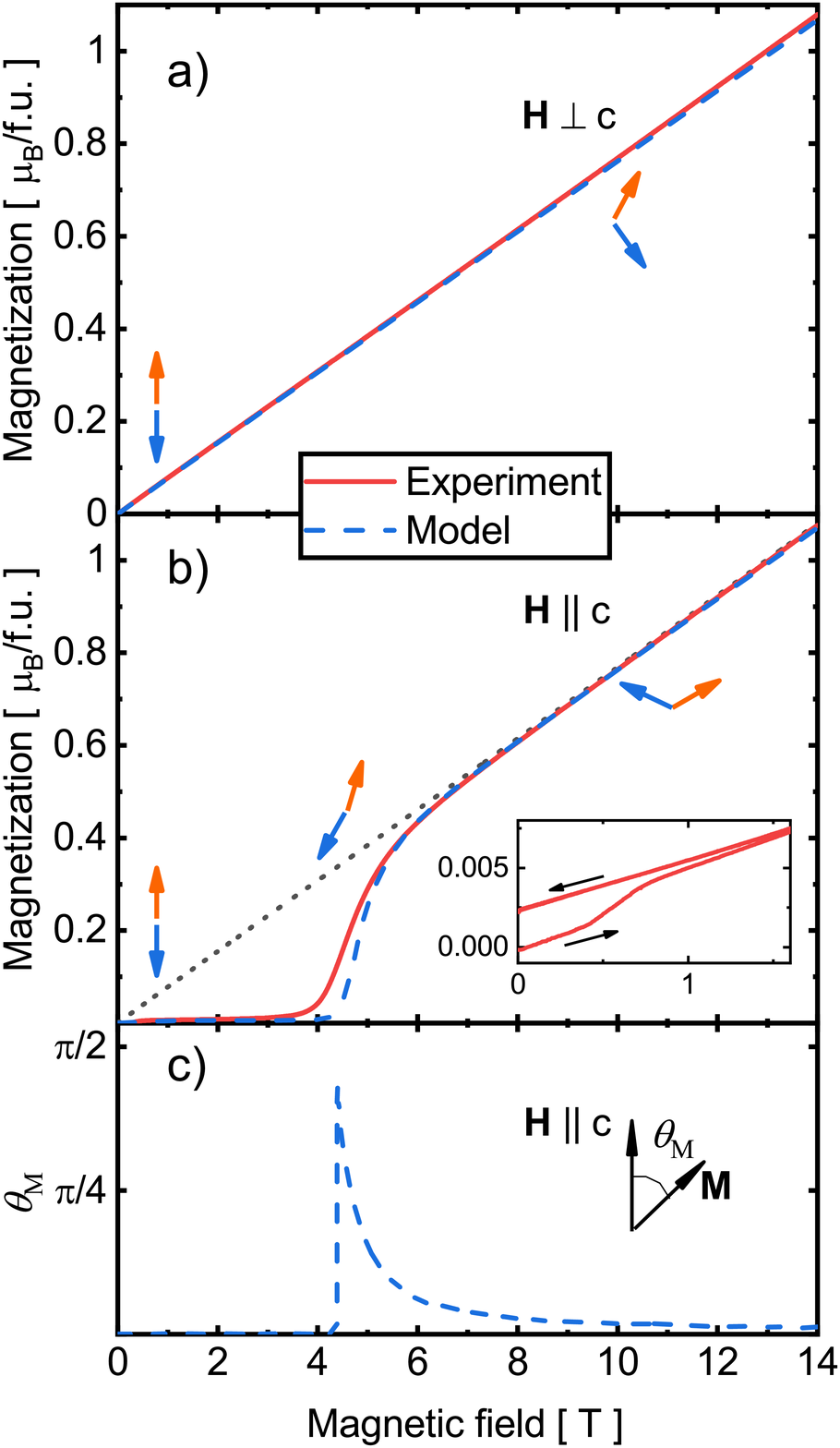}
\end{center}
\vspace{-1cm}
\caption{\textbf{Magnetic field dependence of the magnetization of \Mn~at $T=2\textrm{ K}$.} Magnetic field is perpendicular(a) and parallel(b,c) to the $c$ axis. In all panels the red solid line corresponds to the experiment, while the blue dashed line corresponds to the model calculation. The orange and blue arrows illustrate the magnetic moments of the two sublattices in different magnetic fields pointing horizontally/vertically for (a) and (b), respectively. The experimental curve of the $\vect{H}\perp c$ case is repeated in (b) as black dotted line for comparison. The inset in (b) magnifies the low-field part of the experimental magnetization curve, using the same units as the main axes. Here black arrows indicate the direction of the magnetic field sweep. Panel (c) shows the calculated angle $\theta_\mathrm{M}$ enclosed by the magnetization and the magnetic field.} \label{fig_magn}
\end{figure}

The magnetic field dependence of the magnetization of \Mn~at $T=2\textrm{ K}$ is presented in Fig.~\ref{fig_magn}. While for magnetic fields perpendicular to the hexagonal $c$ axis 
 the magnetization is linear up to $14\textrm{ T}$, for $\vect{H}\parallel c$ the magnetization remains close to zero up to the spin-reorientation transition starting at $\mu_0 H_{C1}=4\textrm{ T}$. In the field range between $\mu_0 H_{C1}$ and $\mu_0 H_{C2}=6\textrm{ T}$ the magnetization smoothly increases, asymptotically reaching the linear susceptibility  of the $\vect{H}\perp c$ case. To complement the magnetization curve presented for a limited magnetic field range in Ref.~[\onlinecite{kurumaji_prb_2017}], the results shown  for a broader magnetic field range in Fig.~\ref{fig_magn}(b) clearly indicate the isotropy of the magnetic susceptibility in fields above $H_{C2}$. 

\begin{figure}[tbp]
\begin{center}
\includegraphics[width=0.99\linewidth, clip]{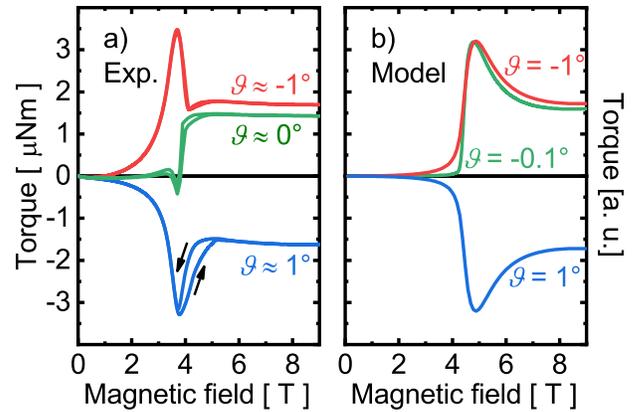}
\end{center}
\vspace{-0.5cm}
\caption{\textbf{Magnetic field dependence of the magnetic torque in \Mn~at $T=2\textrm{ K}$.} (a) experiment and (b) model calculation. The angle included between the $c$ axis and the magnetic field is denoted by $\vartheta$, where the $\pm$ signs represent the directions towards the $\pm a$ axes.} \label{fig_torque}
\end{figure}

The zero magnetization up to $H_{C1}$ for fields along the $c$ axis and the constant susceptibility for the perpendicular direction are characteristic of easy-axis antiferromagnets. The relatively low values of the critical fields along the $c$ axis and the identical susceptibility for $\vect{H}\parallel c$ and $\vect{H}\perp c$ in fields above $H_{C2}$ indicate a nearly isotropic spin-system. The smooth increase of the magnetization between $H_{C1}$ and $H_{C2}$ is a hallmark of competing magnetic anisotropies\cite{turov_book_1965}. In this field region, the magnetization is not parallel to the external field, as shown by the model calculations of the enclosed angle $\theta_M$ in Fig.~\ref{fig_magn}(c) and also evident from the magnetic torque measurements in Fig.~\ref{fig_torque}(a). Namely, torque $(\vect{\tau})$ is produced as the cross product of magnetization $(\vect{M})$ and magnetic field, $\vect{\tau}=\mu_0\vect{M}\times \vect{H}$. When the field is along the $c$ axis, the system is unstable since the direction of the magnetization component perpendicular to the field has no preferred orientation in the $ab$ plane. If the field is rotated by a small $\vartheta$ angle towards the $a$ axis, either clockwise or anticlockwise, a characteristic peak in the magnetic field dependence of the magnetic torque in the $H_{C1}<H<H_{C2}$ field region appears. The small hysteresis, only observed for some angles such as $\vartheta\approx + 1^{\circ}$, and the sign change of the $\vartheta\approx 0^{\circ}$ curve are probably due to the slight mechanical instability of the setup. The approximately saturated torque in high fields is due to the compensation of the decreasing angle $\theta_\mathrm{M}$, Fig.~\ref{fig_magn}(c), by the increasing magnetization amplitude, Fig.~\ref{fig_magn}(b), in increasing field.

\begin{figure*}[tbp]
\begin{center}
\includegraphics[width=0.99\linewidth, clip]{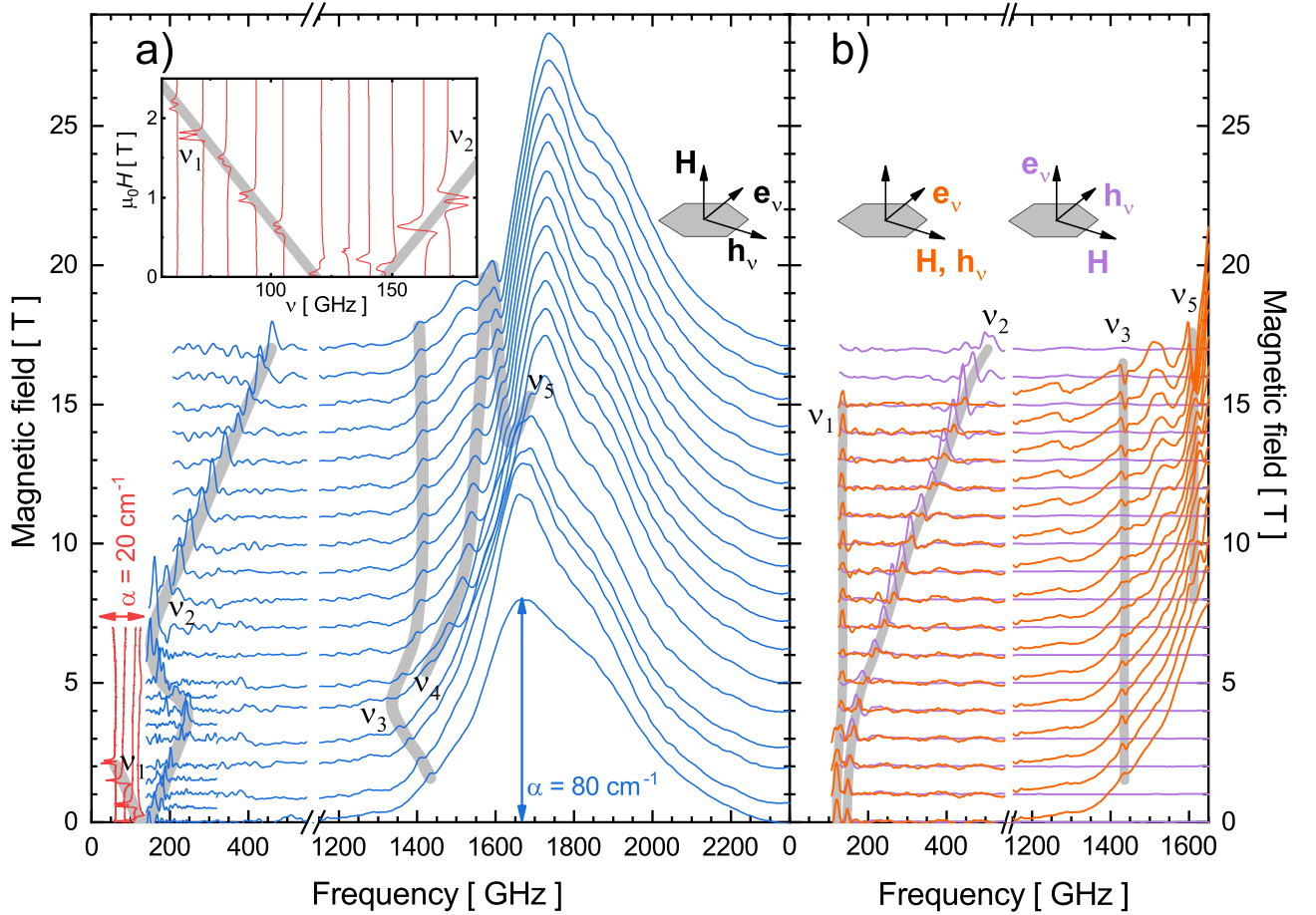}
\end{center}
\vspace{-0.5cm}
\caption{\textbf{Magnetic field dependence of the magnetic absorption in \Mn~at $T=3\textrm{ K}$.} Magnetic field is parallel(a) or perpendicular(b) to the hexagonal $c$ axis. In both panels, absorption spectra are vertically shifted in proportion to the magnetic field. The spectra show the absorption coefficient relative to the paramagnetic phase. Weak resonance modes are marked by light-gray lines which are guides to the eye. The frequency region $550\textrm{ GHz}<\nu<1150\textrm{ GHz}$ between the breakpoints of the horizontal axis is featureless. Magnetic field dependent absorption at fixed low frequencies(a) is shown by vertical red lines, where the baseline is shifted horizontally in proportion to the frequency. Scale of the absorption coefficient for spectra and magnetic field sweeps is indicated by blue/red arrows in (a), respectively. The inset shows a magnified view of the magnetic field dependent absorption at fixed frequencies. In (b), absorption for two perpendicular polarizations is shown in orange ($\vect{e}_\nu,\textrm{ }\vect{h}_\nu\perp c,\textrm{ }\vect{h}_\nu || \vect{H} $) and purple ($\vect{e}_\nu,\textrm{ }\vect{h}_\nu\perp \vect{H},\textrm{ }\vect{e}_\nu || c $), where $\vect{e}_\nu$ and $\vect{h}_\nu$ stand for the electric and magnetic component of light, respectively. In both panels, schematic figures illustrate the measurement geometry, where hexagonal plate shows the $ab$ plane of the crystal.} \label{fig_spectra}
\end{figure*}

To explore the magnetization dynamics in \Mn, we studied the magnetic field dependence of the magnetic resonances using THz absorption spectroscopy at low temperatures $(T=3\textrm{ K})$. 
The lowest-frequency resonances $(\nu<150\textrm{ GHz})$, which are not accessible by far-infrared spectroscopy, were investigated by the use of backward-wave oscillators.  Although the absorption line shapes were distorted by diffraction on the edges of the sample, and by interference effects, as seen in the inset of Fig.~\ref{fig_spectra}(a), the resonance frequencies and selection rules could be determined with reasonable accuracy.         

\begin{figure*}[tbp]
\begin{center}
\includegraphics[width=0.99\linewidth, clip]{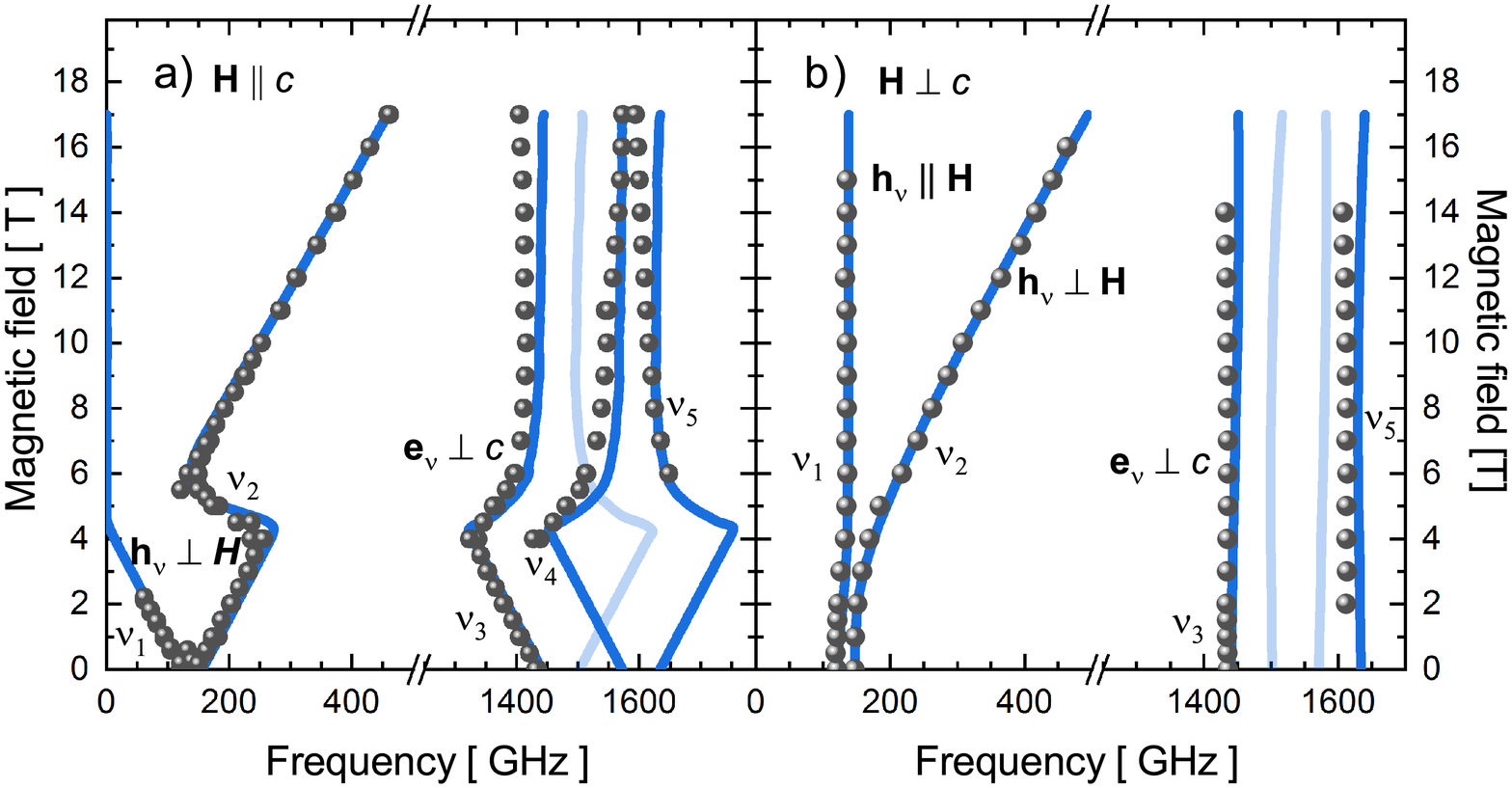}
\end{center}
\vspace{-0.5cm}
\caption{\textbf{Magnetic field dependence of the magnetic resonance frequencies of \Mn~at $T=3\textrm{ K}$ as seen by far-infrared and microwave optical transmission spectroscopy.} Magnetic field is parallel(a) or perpendicular(b) to the crystallographic $c$ axis. In both panels black spheres correspond to the experimental values, while blue lines to the model calculation result. Resonance modes predicted by theory but not active in the optical experiments are shown in light blue. Observed optical selection rules are indicated next to the corresponding excitation branches.} \label{fig_freq}
\end{figure*}

Typical absorption spectra, presented in Fig.~\ref{fig_spectra}(a), contain a strong absorption band between 1400\,GHz and 2200\,GHz and additional weaker absorption peaks. We performed a polarization-dependent study in order to clarify the selection rules of the modes. While in the low-frequency region $(\nu<500\textrm{ GHz})$ the modes are sensitive to the magnetic component of light (details shown in Fig.~\ref{fig_freq} and discussed later), the high-frequency modes, $\nu>1200\textrm{ GHz}$, are electric dipole active and they can be excited by an oscillating electric field perpendicular to the $c$ axis. In the following, we focus on the field dependence of the weak resonance modes 
associated with magnon excitations and leave the strong absorption band between $1400\textrm{ GHz}<\nu<2200\textrm{ GHz}$ to later studies. Moreover, temperature-dependent measurements show that this broad excitation is not restricted to the magnetically ordered state, though it shows some field- and temperature dependence. 

In zero magnetic field, we observed three resonance branches, $\nu_{1}$, $\nu_{2}$, and $\nu_{3}$, as presented in the mode map of Fig.~\ref{fig_freq}(a). While the lower-frequency modes $\nu_{1}$ and $\nu_{2}$ are magnetic dipole active, the high-frequency $\nu_{3}$ shows electric dipole activity along the $c$ axis. Starting from zero magnetic field, in increasing fields parallel to the easy-axis ($\vect{H}\parallel c$, Fig.~\ref{fig_freq}(a))  $\nu_{1}$ and  $\nu_{3}$ shift towards lower frequencies, whilst $\nu_{2}$ moves to higher frequencies. Within the experimental precision, in all three cases the slope of the shift corresponds to the free electron spin $g$-factor of $g_e=2$\cite{odom_prl_2007}. In finite fields, $\nu_{1}$ and  $\nu_{2}$ are excited by an oscillating magnetic field, $\vect{h}_\nu$, perpendicular to the static one, $\vect{h}_\nu \perp \vect{H}$.  Between $\mu_0 H_{C1}=4\textrm{ T}$ and $\mu_0 H_{C2}=6\textrm{ T}$, the slopes of the resonances $\nu_{2}$ and $\nu_{3}$ change sign and one can observe an additional electric dipole active mode, $\nu_{4}$, with a positive slope at high frequencies. Above $H_{C2}$, the $\nu_{2}$ mode shifts towards higher frequencies again, with a slope corresponding to $g_e$, while the frequencies $\nu_{3}$ and $\nu_{4}$ remain roughly constant in increasing fields. In this phase a new electric dipole active mode, $\nu_{5}$, with almost field-independent frequency appears. The fact that the $g$-factors of the modes are identical, further supports the nearly isotropic nature of the spin system. 

Considering the smooth increase of the magnetization in the $H_{C1}<H<H_{C2}$ phase transition region (Fig.~\ref{fig_magn}(b)), without hysteresis, the phase transition can be categorized as of second order. This classification is further supported by the torque curves (Fig.~\ref{fig_torque}(a), green and red) being also free of hysteresis. Moreover, by fitting the linear field dependence of the $\nu_{1}$ resonance frequency, one can conclude that it softens to zero frequency at the $H_{C1}$ critical field, characteristic for second order transitions.

In magnetic fields perpendicular to $c$, see Fig.~\ref{fig_spectra}(b), $\nu_{1}$ is active for $\vect{h}_\nu \parallel \vect{H}$ and its frequency shows almost no field dependence. In contrast, $\nu_{2}$ is excited by $\vect{h}_\nu \perp \vect{H}$ and shifts towards higher frequencies with increasing field, with the slope corresponding to $g_e$. We could not observe the weak $\nu_{4}$ resonance in the $\vect{H} \perp c$ geometry, while $\nu_{3}$ and $\nu_{5}$ are active for oscillating electric fields, $\vect{e}_\nu \perp c$ and their frequency is independent of the magnetic field. 

Next, we turn to the modelling of the spin system. The magnetization and the spin resonances of \Mn~are consistent with a relatively simple, two-sublattice ferrimagnetic model, that we discuss in the following. The two sublattices interact antiferromagnetically and experience different weak single-ion magnetic anisotropies that can be attributed to the octahedral/tetrahedral enviroments, respectively. 
The $g$-factors at the two crystallographic sites can also be different.
The corresponding Hamiltonian in the mean-field approach is
\begin{eqnarray}
\label{Eq_ham}
\mathcal{H}&=&J\vect{S}_1\vect{S}_2 + \Delta_1\left(S_1^c\right)^2 + \Delta_2\left(S_2^c\right)^2 \nonumber\\
&-& \mu_B\mu_0\vect{H}\left(g_1\vect{S}_1 + g_2\vect{S}_2\right),
\end{eqnarray}
where $\vect{S}_1$ and $\vect{S}_2$ are the three-dimensional vectors of $S=5/2$ lengths representing the magnetic moments of the two sublattices. In the Hamiltonian the dominant energy scale is given by the antiferromagnetic coupling,  $J$, while the $\Delta_1$ and $\Delta_2$ single-ion $c$-axis anisotropies define the zero-field spin orientation. In the Zeeman term $\mu_B$ is the Bohr magneton, $g_{1}$ and $g_{2}$ are the $g$-factors of the two sublattices, and $\mu_0$ denotes the permeability of vacuum.  

The ground state at a given external field value can be determined by minimizing the energy given in Eq.~\ref{Eq_ham}. The easy-axis collinear state is realized in zero field if $\Delta_1 + \Delta_2<0$. To reproduce the smooth increase of magnetization in the $\vect{H}\parallel c$, $H_{C1}<H<H_{C2}$ case, $\Delta_1$ and $\Delta_2$ have to have opposite signs.\cite{turov_book_1965} In this field range the magnetic moments of the two sublattices continuously rotate to become almost perpendicular to the increasing magnetic field. Since $\Delta_1$ and $\Delta_2$  are different, the magnetization is not parallel to the applied field, as clear from Fig.~\ref{fig_magn}(c), which explains the peak of the magnetic torque in the $H_{C1}<H<H_{C2}$ field range, as presented in Fig.~\ref{fig_torque}. 


Within the model described by Eq.~\ref{Eq_ham} the isotropic high-field $(H>H_{C2})$ differential susceptibility is
\begin{equation}
\label{chi}
\chi=\frac{(g_1 + g_2)^2}{4J-2(\Delta_1+\Delta_2)},
\end{equation}
if the second-order terms in the difference of the two sublattices, $\sim\left(\Delta_1 - \Delta_2\right)^2$ and $\sim\left(g_1 - g_2\right)^2$, are neglected. From the lowest zero-field resonance frequencies, $\nu_{1}$ and $\nu_{2}$, the model parameters $J$, $\Delta_1$, and $\Delta_2$ can be determined, according to
\begin{equation}
\label{nu}
\nu_{1,2}=\left(\sqrt{-2 J (\Delta_1+\Delta_2) + (\Delta_1+\Delta_2)^2} \pm (\Delta_1-\Delta_2)\right)S\textrm{.}
\end{equation}
 The slopes of $\nu_{1}$ and $\nu_{2}$ in $\vect{H}\parallel c$, $H<H_{C1}$ correspond to the $g$-factors $g_{1}\approx g_{2}\approx  2$. The zero-field remanent magnetization at $T=2\textrm{ K}$ after a $\vect{H}\parallel c$ field treatment, $M_0=0.0025~\mu_B/\textrm{f.u.}$ as found in Ref.~[\onlinecite{kurumaji_prb_2017}] and in our experiments shown in the inset of Fig.~\ref{fig_magn}(b), is a consequence of the different $g$-factors of the two sublattices, thus $M_0=(g_1-g_2)S$, since sublattice 1 with the slightly larger $g$-factor aligns along the field, while sublattice 2 turns opposite to it. 

From the considerations above, the field dependence of the magnetization, the torque and the magnon modes can all be well reproduced by the following parameter set: $J=3\textrm{ meV}$, $\Delta_1=-0.015\textrm{ meV}$, $\Delta_2=0.006\textrm{ meV}$, $g_1=2.001$, and $g_2=2$. Using these parameters, we numerically found the minimum of the energy in Eq.~\ref{Eq_ham} at various magnetic fields. The calculated magnetization of the ground state is presented in Figs.~\ref{fig_magn}(a,b) by blue lines, while Fig.~\ref{fig_torque}(b) shows the calculated magnetic torque, both in a good correspondence with the experiments. However, the lower critical field, $H_{C1}$ is slightly overestimated by the model. Since the width of the peak in the field dependence of the torque is proportional to the $H_{C1}$ field\cite{torque}, the model results a broader peak in Fig.~\ref{fig_torque}(b) than the experiments (Fig.~\ref{fig_torque}(a)).   

The model parameters are independently confirmed by \textit{ab initio} calculations that return $J=2.7$\,meV, $g_1=g_2=2.002$, $\Delta_1=-0.0015$\,meV, and $\Delta_2=0.001$\,meV, where easy-axis ($\Delta_1<0$) and easy-plane ($\Delta_2>0$) anisotropies are obtained at the octahedral and tetrahedral sites, respectively. The absolute values of $\Delta_1$ and $\Delta_2$ are underestimated, though, which may be due to the systematic error of DFT. Nevertheless, with the help of DFT one can identify the sublattice $1$ and $2$ with the octahedrally and tetrahedrally coordinated Mn sites. 

The magnetic resonances were determined by calculating the response to small perturbations in the ground state, as descibed in detail in Refs.~[\onlinecite{szaller_prb_2017}] and [\onlinecite{peedu_prb_2019}]. The results are shown in Figs.~\ref{fig_freq}(a,b). The two ferrimagnetically ordered classical spins have two $\Gamma$-point (zero momentum) magnon excitations, which correspond to $\nu_{1}$ and $\nu_{2}$. For $\vect{H}\parallel c$, these two modes are excited by circularly polarized light of opposite helicity propagating along the direction of the magnetic field, thus in the linearly polarized optical experiments they are visible in the $\vect{h}_\nu \perp \vect{H}$ geometry. On the other hand, for sufficiently high magnetic field $\vect{H}\perp c$ the $\nu_{1}$ excitation corresponds to the quasi-antiferromagnetic while $\nu_{2}$ to the quasi-ferromagnetic resonance, namely, $\nu_{1}$ and $\nu_{2}$ correspond to the precession of $\vect{L}=\vect{S}_1-\vect{S}_2$ and $\vect{M}=\vect{S}_1+\vect{S}_2$, respectively. Accordingly, the selection rules are $\vect{h}_\nu \parallel \vect{H}$ for $\nu_{1}$ and $\vect{h}_\nu \perp \vect{H}$ for $\nu_{2}$, both as in the experiments and in the model.  

\begin{figure}[tbp]
\begin{center}
\includegraphics[width=0.99\linewidth, clip]{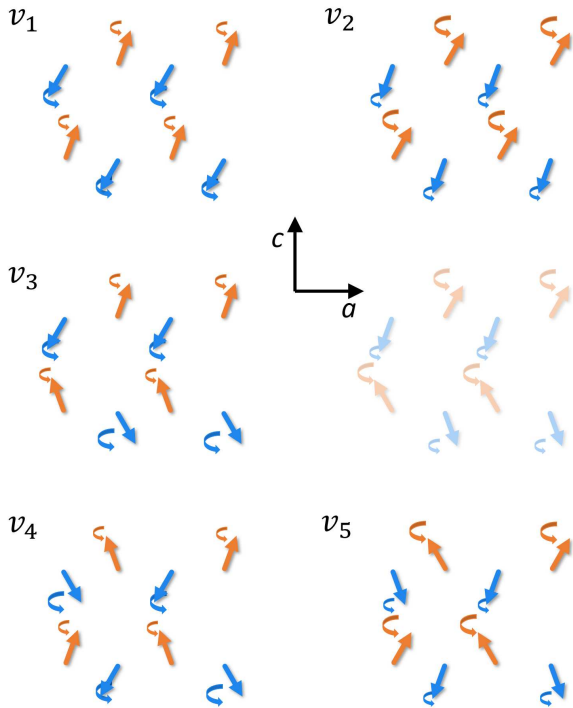}
\end{center}
\vspace{-0.5cm}
\caption{\textbf{Magnetic resonance modes of \Mn.} Orange and blue vectors, corresponding to the octa- and tetrahedrally coordinated Mn spins, show snapshots of each excitation mode in $H<H_{C1}$ magnetic field along the positive $c$ axis. The traces of the tips of all spins describe circles in the $ab$ plane of the crystal. The relative amplitudes are indicated by the size of the curved arrows. The mode corresponding to the faint cartoon was not observed in the experiments.} \label{fig_modes}
\end{figure}

The electric dipole active high-frequency $(1200\textrm{ GHz}<\nu)$ modes, $\nu_{3}$, $\nu_{4}$ and $\nu_{5}$, can be interpreted as Brillouin zone-edge magnons of the simplified model system with two spins in the unit cell. However, these resonances correspond to $\Gamma$-point excitations if we consider that the crystallographic unit cell contains four Mn sites, as shown in Fig.~\ref{fig_struct}. With the assumption of a four-spin magnetic unit cell, one can extend the model in Eq.~\ref{Eq_ham} to distinguish between first-neighbour antiferromagnetic interactions of octahedrally and tetrahedrally coordinated spins within the $ab$ plane, $J_1$, and between adjacent $ab$ planes, $J_2$. To numerically reproduce the zero-field splitting of the $\nu_{4}$ and $\nu_{5}$ resonances, we also consider a weak antiferromagnetic $J_{3}$ coupling between tetrahedrally coordinated spins of neighbouring $ab$ planes, as presented in Fig.~\ref{fig_struct}. To ensure the compatibility with the previous results of Eq.~\ref{Eq_ham}, $J=3J_1 + 2 J_2$ has to hold, where the integers correspond to the coordination numbers. In this context, $\nu_{3}$, $\nu_{4}$ and $\nu_{5}$ can be viewed as excitations with $\pi$ phase shifts between the equivalently coordinated spins of adjacent $ab$ planes. For $\nu_{3}$ the crystallographically equivalent spins oscillate in-phase within a single $ab$ plane, while for $\nu_{4}$ and $\nu_{5}$ four of their closest-neighbours have opposite, and two of them the same phase, as illustrated in Fig.~\ref{fig_modes}. Within this extended model, using $J_1=0.8\textrm{ meV}$, $J_2=0.31\textrm{ meV}$ and $J_{3}=0.005\textrm{ meV}$ values, the magnetic field dependence of the $\nu_{3}$, $\nu_{4}$ and $\nu_{5}$ resonances is reproduced both in the  $\vect{H}\parallel c$ and  $\vect{H}\perp c$ geometries, as presented in Figs.~\ref{fig_freq}(a,b). To explain their $\vect{e}_\nu \perp c$ optical selection rule, one has to consider a model including spin-polarization coupling where the electric dipole moment of these resonances can be calculated, which is out of the scope of the present study.  


In conclusion, the low-temperature static magnetic properties and spin excitations of \Mn~were investigated by various experimental techniques. The observed magnetic field dependences of the magnetization, torque, and spin-wave resonance frequencies are reproduced by a mean-field model. The magnetic exchange and $g$-factor parameters of the model were determined by fitting the rich experimental dataset, and are also supported by first principle calculations. The quantitative explanation of the various magnetic resonances of \Mn~can serve as a starting point to the understanding of the more complicated excitations of other compounds in the $M_2$Mo$_3$O$_8$ family\cite{kurumaji_prl_2017,kurumaji_prb_2017}.

\subsection*{Acknowledgements}
This project was supported by institutional research funding IUT23-3 of the Estonian Ministry of Education and Research, by the European Regional Development Fund project TK134, by the bilateral program of the Estonian and Hungarian Academies of Sciences grant NKM2018-47, by the Hungarian NKFIH grants ANN 122879 and 2019-2.1.11-TÉT-2019-00029, by the Austrian Agency for International Cooperation in Education and Research grant WTZ HU 08/2020, by the Austrian Science Funds grant I 2816-N27, and by the grant ANCD 20.80009.5007.19 (Rep. of Moldova). This work was partly supported by the Deutsche Forschungsgemeinschaft(DFG) through grant No. JE748/1 and Transregional Research Collaboration TRR 80 (Augsburg, Munich, and Stuttgart).


\begin{thebibliography}{55}%
\makeatletter
\providecommand \@ifxundefined [1]{%
 \@ifx{#1\undefined}
}%
\providecommand \@ifnum [1]{%
 \ifnum #1\expandafter \@firstoftwo
 \else \expandafter \@secondoftwo
 \fi
}%
\providecommand \@ifx [1]{%
 \ifx #1\expandafter \@firstoftwo
 \else \expandafter \@secondoftwo
 \fi
}%
\providecommand \natexlab [1]{#1}%
\providecommand \enquote  [1]{``#1''}%
\providecommand \bibnamefont  [1]{#1}%
\providecommand \bibfnamefont [1]{#1}%
\providecommand \citenamefont [1]{#1}%
\providecommand \href@noop [0]{\@secondoftwo}%
\providecommand \href [0]{\begingroup \@sanitize@url \@href}%
\providecommand \@href[1]{\@@startlink{#1}\@@href}%
\providecommand \@@href[1]{\endgroup#1\@@endlink}%
\providecommand \@sanitize@url [0]{\catcode `\\12\catcode `\$12\catcode
  `\&12\catcode `\#12\catcode `\^12\catcode `\_12\catcode `\%12\relax}%
\providecommand \@@startlink[1]{}%
\providecommand \@@endlink[0]{}%
\providecommand \url  [0]{\begingroup\@sanitize@url \@url }%
\providecommand \@url [1]{\endgroup\@href {#1}{\urlprefix }}%
\providecommand \urlprefix  [0]{URL }%
\providecommand \Eprint [0]{\href }%
\providecommand \doibase [0]{http://dx.doi.org/}%
\providecommand \selectlanguage [0]{\@gobble}%
\providecommand \bibinfo  [0]{\@secondoftwo}%
\providecommand \bibfield  [0]{\@secondoftwo}%
\providecommand \translation [1]{[#1]}%
\providecommand \BibitemOpen [0]{}%
\providecommand \bibitemStop [0]{}%
\providecommand \bibitemNoStop [0]{.\EOS\space}%
\providecommand \EOS [0]{\spacefactor3000\relax}%
\providecommand \BibitemShut  [1]{\csname bibitem#1\endcsname}%
\let\auto@bib@innerbib\@empty
\bibitem [{\citenamefont {Kimura}\ \emph {et~al.}(2003)\citenamefont {Kimura},
  \citenamefont {Goto}, \citenamefont {Shintani}, \citenamefont {Ishizaka},
  \citenamefont {Arima},\ and\ \citenamefont {Tokura}}]{kimura_nature_2003}%
  \BibitemOpen
  \bibfield  {author} {\bibinfo {author} {\bibfnamefont {T.}~\bibnamefont
  {Kimura}}, \bibinfo {author} {\bibfnamefont {T.}~\bibnamefont {Goto}},
  \bibinfo {author} {\bibfnamefont {H.}~\bibnamefont {Shintani}}, \bibinfo
  {author} {\bibfnamefont {K.}~\bibnamefont {Ishizaka}}, \bibinfo {author}
  {\bibfnamefont {T.}~\bibnamefont {Arima}}, \ and\ \bibinfo {author}
  {\bibfnamefont {Y.}~\bibnamefont {Tokura}},\ }\bibfield  {title} {\enquote
  {\bibinfo {title} {Magnetic control of ferroelectric polarization},}\ }\href
  {\doibase 10.1038/nature02018} {\bibfield  {journal} {\bibinfo  {journal}
  {Nature (London)}\ }\textbf {\bibinfo {volume} {426}},\ \bibinfo {pages} {55}
  (\bibinfo {year} {2003})}\BibitemShut {NoStop}%
\bibitem [{\citenamefont {Fiebig}(2005)}]{fiebig_jpd_2005}%
  \BibitemOpen
  \bibfield  {author} {\bibinfo {author} {\bibfnamefont {M.}~\bibnamefont
  {Fiebig}},\ }\bibfield  {title} {\enquote {\bibinfo {title} {Revival of the
  magnetoelectric effect},}\ }\href {http://stacks.iop.org/0022-3727/38/R123}
  {\bibfield  {journal} {\bibinfo  {journal} {J. Phys. D: Appl. Phys.}\
  }\textbf {\bibinfo {volume} {38}},\ \bibinfo {pages} {R123} (\bibinfo {year}
  {2005})}\BibitemShut {NoStop}%
\bibitem [{\citenamefont {Spaldin}\ and\ \citenamefont
  {Fiebig}(2005)}]{spaldin_science_2005}%
  \BibitemOpen
  \bibfield  {author} {\bibinfo {author} {\bibfnamefont {Nicola~A}\
  \bibnamefont {Spaldin}}\ and\ \bibinfo {author} {\bibfnamefont {Manfred}\
  \bibnamefont {Fiebig}},\ }\bibfield  {title} {\enquote {\bibinfo {title} {The
  renaissance of magnetoelectric multiferroics},}\ }\href@noop {} {\bibfield
  {journal} {\bibinfo  {journal} {Science}\ }\textbf {\bibinfo {volume}
  {309}},\ \bibinfo {pages} {391--392} (\bibinfo {year} {2005})}\BibitemShut
  {NoStop}%
\bibitem [{\citenamefont {Eerenstein}\ \emph {et~al.}(2006)\citenamefont
  {Eerenstein}, \citenamefont {Mathur},\ and\ \citenamefont
  {Scott}}]{eerenstein_nature_2006}%
  \BibitemOpen
  \bibfield  {author} {\bibinfo {author} {\bibfnamefont {W.}~\bibnamefont
  {Eerenstein}}, \bibinfo {author} {\bibfnamefont {N.~D.}\ \bibnamefont
  {Mathur}}, \ and\ \bibinfo {author} {\bibfnamefont {J.~F.}\ \bibnamefont
  {Scott}},\ }\bibfield  {title} {\enquote {\bibinfo {title} {Multiferroic and
  magnetoelectric materials},}\ }\href {\doibase 10.1038/nature05023}
  {\bibfield  {journal} {\bibinfo  {journal} {Nature (London)}\ }\textbf
  {\bibinfo {volume} {442}},\ \bibinfo {pages} {759} (\bibinfo {year}
  {2006})}\BibitemShut {NoStop}%
\bibitem [{\citenamefont {Cheong}\ and\ \citenamefont
  {Mostovoy}(2007)}]{cheong_nmat_2007}%
  \BibitemOpen
  \bibfield  {author} {\bibinfo {author} {\bibfnamefont {S.-W.}\ \bibnamefont
  {Cheong}}\ and\ \bibinfo {author} {\bibfnamefont {M.}~\bibnamefont
  {Mostovoy}},\ }\bibfield  {title} {\enquote {\bibinfo {title} {Multiferroics:
  a magnetic twist for ferroelectricity},}\ }\href {\doibase 10.1038/nmat1804}
  {\bibfield  {journal} {\bibinfo  {journal} {Nat. Mater.}\ }\textbf {\bibinfo
  {volume} {6}},\ \bibinfo {pages} {13} (\bibinfo {year} {2007})}\BibitemShut
  {NoStop}%
\bibitem [{\citenamefont {Wu}\ \emph {et~al.}(2013)\citenamefont {Wu},
  \citenamefont {Cybart}, \citenamefont {Yi}, \citenamefont {Parker},
  \citenamefont {Ramesh},\ and\ \citenamefont {Dynes}}]{wu_prl_2013}%
  \BibitemOpen
  \bibfield  {author} {\bibinfo {author} {\bibfnamefont {S.~M.}\ \bibnamefont
  {Wu}}, \bibinfo {author} {\bibfnamefont {Shane~A.}\ \bibnamefont {Cybart}},
  \bibinfo {author} {\bibfnamefont {D.}~\bibnamefont {Yi}}, \bibinfo {author}
  {\bibfnamefont {James~M.}\ \bibnamefont {Parker}}, \bibinfo {author}
  {\bibfnamefont {R.}~\bibnamefont {Ramesh}}, \ and\ \bibinfo {author}
  {\bibfnamefont {R.~C.}\ \bibnamefont {Dynes}},\ }\bibfield  {title} {\enquote
  {\bibinfo {title} {Full electric control of exchange bias},}\ }\href
  {\doibase 10.1103/PhysRevLett.110.067202} {\bibfield  {journal} {\bibinfo
  {journal} {Phys. Rev. Lett.}\ }\textbf {\bibinfo {volume} {110}},\ \bibinfo
  {pages} {067202} (\bibinfo {year} {2013})}\BibitemShut {NoStop}%
\bibitem [{\citenamefont {Dong}\ \emph {et~al.}(2015)\citenamefont {Dong},
  \citenamefont {Liu}, \citenamefont {Cheong},\ and\ \citenamefont
  {Ren}}]{dong_advph_2015}%
  \BibitemOpen
  \bibfield  {author} {\bibinfo {author} {\bibfnamefont {Sh.}\ \bibnamefont
  {Dong}}, \bibinfo {author} {\bibfnamefont {J.-M.}\ \bibnamefont {Liu}},
  \bibinfo {author} {\bibfnamefont {S.-W.}\ \bibnamefont {Cheong}}, \ and\
  \bibinfo {author} {\bibfnamefont {Zh.}\ \bibnamefont {Ren}},\ }\bibfield
  {title} {\enquote {\bibinfo {title} {Multiferroic materials and
  magnetoelectric physics: symmetry, entanglement, excitation, and topology},}\
  }\href {\doibase 10.1080/00018732.2015.1114338} {\bibfield  {journal}
  {\bibinfo  {journal} {Adv. Phys.}\ }\textbf {\bibinfo {volume} {64}},\
  \bibinfo {pages} {519--626} (\bibinfo {year} {2015})}\BibitemShut {NoStop}%
\bibitem [{\citenamefont {Fiebig}\ \emph {et~al.}(2016)\citenamefont {Fiebig},
  \citenamefont {Lottermoser}, \citenamefont {Meier},\ and\ \citenamefont
  {Trassin}}]{fiebig_natrev_2016}%
  \BibitemOpen
  \bibfield  {author} {\bibinfo {author} {\bibfnamefont {M.}~\bibnamefont
  {Fiebig}}, \bibinfo {author} {\bibfnamefont {Th.}\ \bibnamefont
  {Lottermoser}}, \bibinfo {author} {\bibfnamefont {D.}~\bibnamefont {Meier}},
  \ and\ \bibinfo {author} {\bibfnamefont {M.}~\bibnamefont {Trassin}},\
  }\bibfield  {title} {\enquote {\bibinfo {title} {The evolution of
  multiferroics},}\ }\href {\doibase 10.1038/natrevmats.2016.46} {\bibfield
  {journal} {\bibinfo  {journal} {Nat. Rev. Mater.}\ }\textbf {\bibinfo
  {volume} {1}},\ \bibinfo {pages} {16046} (\bibinfo {year}
  {2016})}\BibitemShut {NoStop}%
\bibitem [{\citenamefont {Kuzmenko}\ \emph {et~al.}(2018)\citenamefont
  {Kuzmenko}, \citenamefont {Szaller}, \citenamefont {Kain}, \citenamefont
  {Dziom}, \citenamefont {Weymann}, \citenamefont {Shuvaev}, \citenamefont
  {Pimenov}, \citenamefont {Mukhin}, \citenamefont {Ivanov}, \citenamefont
  {Gudim}, \citenamefont {Bezmaternykh},\ and\ \citenamefont
  {Pimenov}}]{kuzmenko_prl_2018}%
  \BibitemOpen
  \bibfield  {author} {\bibinfo {author} {\bibfnamefont {A.~M.}\ \bibnamefont
  {Kuzmenko}}, \bibinfo {author} {\bibfnamefont {D.}~\bibnamefont {Szaller}},
  \bibinfo {author} {\bibfnamefont {Th.}\ \bibnamefont {Kain}}, \bibinfo
  {author} {\bibfnamefont {V.}~\bibnamefont {Dziom}}, \bibinfo {author}
  {\bibfnamefont {L.}~\bibnamefont {Weymann}}, \bibinfo {author} {\bibfnamefont
  {A.}~\bibnamefont {Shuvaev}}, \bibinfo {author} {\bibfnamefont {Anna}\
  \bibnamefont {Pimenov}}, \bibinfo {author} {\bibfnamefont {A.~A.}\
  \bibnamefont {Mukhin}}, \bibinfo {author} {\bibfnamefont {V.~Yu.}\
  \bibnamefont {Ivanov}}, \bibinfo {author} {\bibfnamefont {I.~A.}\
  \bibnamefont {Gudim}}, \bibinfo {author} {\bibfnamefont {L.~N.}\ \bibnamefont
  {Bezmaternykh}}, \ and\ \bibinfo {author} {\bibfnamefont {A.}~\bibnamefont
  {Pimenov}},\ }\bibfield  {title} {\enquote {\bibinfo {title} {Switching of
  magnons by electric and magnetic fields in multiferroic borates},}\ }\href
  {\doibase 10.1103/PhysRevLett.120.027203} {\bibfield  {journal} {\bibinfo
  {journal} {Phys. Rev. Lett.}\ }\textbf {\bibinfo {volume} {120}},\ \bibinfo
  {pages} {027203} (\bibinfo {year} {2018})}\BibitemShut {NoStop}%
\bibitem [{\citenamefont {Weymann}\ \emph {et~al.}(2020)\citenamefont
  {Weymann}, \citenamefont {Bergen}, \citenamefont {Kain}, \citenamefont
  {Pimenov}, \citenamefont {Shuvaev}, \citenamefont {Constable}, \citenamefont
  {Szaller}, \citenamefont {Mill}, \citenamefont {Kuzmenko}, \citenamefont
  {Ivanov}, \citenamefont {Kostyuchenko}, \citenamefont {Popov}, \citenamefont
  {Zvezdin}, \citenamefont {Pimenov}, \citenamefont {Mukhin},\ and\
  \citenamefont {Mostovoy}}]{weymann_npjqm_2020}%
  \BibitemOpen
  \bibfield  {author} {\bibinfo {author} {\bibfnamefont {Lukas}\ \bibnamefont
  {Weymann}}, \bibinfo {author} {\bibfnamefont {Lorenz}\ \bibnamefont
  {Bergen}}, \bibinfo {author} {\bibfnamefont {Thomas}\ \bibnamefont {Kain}},
  \bibinfo {author} {\bibfnamefont {Anna}\ \bibnamefont {Pimenov}}, \bibinfo
  {author} {\bibfnamefont {Alexey}\ \bibnamefont {Shuvaev}}, \bibinfo {author}
  {\bibfnamefont {Evan}\ \bibnamefont {Constable}}, \bibinfo {author}
  {\bibfnamefont {David}\ \bibnamefont {Szaller}}, \bibinfo {author}
  {\bibfnamefont {Boris~V.}\ \bibnamefont {Mill}}, \bibinfo {author}
  {\bibfnamefont {Artem~M.}\ \bibnamefont {Kuzmenko}}, \bibinfo {author}
  {\bibfnamefont {Vsevolod~Yu.}\ \bibnamefont {Ivanov}}, \bibinfo {author}
  {\bibfnamefont {Nadezhda~V.}\ \bibnamefont {Kostyuchenko}}, \bibinfo {author}
  {\bibfnamefont {Alexander~I.}\ \bibnamefont {Popov}}, \bibinfo {author}
  {\bibfnamefont {Anatoly~K.}\ \bibnamefont {Zvezdin}}, \bibinfo {author}
  {\bibfnamefont {Andrei}\ \bibnamefont {Pimenov}}, \bibinfo {author}
  {\bibfnamefont {Alexander~A.}\ \bibnamefont {Mukhin}}, \ and\ \bibinfo
  {author} {\bibfnamefont {Maxim}\ \bibnamefont {Mostovoy}},\ }\bibfield
  {title} {\enquote {\bibinfo {title} {Unusual magnetoelectric effect in
  paramagnetic rare-earth langasite},}\ }\href {\doibase
  10.1038/s41535-020-00263-9} {\bibfield  {journal} {\bibinfo  {journal} {npj
  Quantum Materials}\ }\textbf {\bibinfo {volume} {5}},\ \bibinfo {pages} {61}
  (\bibinfo {year} {2020})}\BibitemShut {NoStop}%
\bibitem [{\citenamefont {Szaller}\ \emph {et~al.}(2019)\citenamefont
  {Szaller}, \citenamefont {Shuvaev}, \citenamefont {Mukhin}, \citenamefont
  {Kuzmenko},\ and\ \citenamefont {Pimenov}}]{szaller_psr_2019}%
  \BibitemOpen
  \bibfield  {author} {\bibinfo {author} {\bibfnamefont {D.}~\bibnamefont
  {Szaller}}, \bibinfo {author} {\bibfnamefont {A.}~\bibnamefont {Shuvaev}},
  \bibinfo {author} {\bibfnamefont {A.~A.}\ \bibnamefont {Mukhin}}, \bibinfo
  {author} {\bibfnamefont {A.~M.}\ \bibnamefont {Kuzmenko}}, \ and\ \bibinfo
  {author} {\bibfnamefont {A.}~\bibnamefont {Pimenov}},\ }\bibfield  {title}
  {\enquote {\bibinfo {title} {Controlling of light with electromagnons},}\
  }\href {\doibase 10.1515/psr-2019-0055} {\bibfield  {journal} {\bibinfo
  {journal} {Physical Sciences Reviews}\ }\textbf {\bibinfo {volume} {5}},\
  \bibinfo {pages} {0055} (\bibinfo {year} {2019})}\BibitemShut {NoStop}%
\bibitem [{\citenamefont {K\'ezsm\'arki}\ \emph {et~al.}(2011)\citenamefont
  {K\'ezsm\'arki}, \citenamefont {Kida}, \citenamefont {Murakawa},
  \citenamefont {Bord\'acs}, \citenamefont {Onose},\ and\ \citenamefont
  {Tokura}}]{kezsmarki_prl_2011}%
  \BibitemOpen
  \bibfield  {author} {\bibinfo {author} {\bibfnamefont {I.}~\bibnamefont
  {K\'ezsm\'arki}}, \bibinfo {author} {\bibfnamefont {N.}~\bibnamefont {Kida}},
  \bibinfo {author} {\bibfnamefont {H.}~\bibnamefont {Murakawa}}, \bibinfo
  {author} {\bibfnamefont {S.}~\bibnamefont {Bord\'acs}}, \bibinfo {author}
  {\bibfnamefont {Y.}~\bibnamefont {Onose}}, \ and\ \bibinfo {author}
  {\bibfnamefont {Y.}~\bibnamefont {Tokura}},\ }\bibfield  {title} {\enquote
  {\bibinfo {title} {Enhanced directional dichroism of terahertz light in
  resonance with magnetic excitations of the multiferroic
  {Ba}$_2${CoGe}$_2${O}$_7$ oxide compound},}\ }\href {\doibase
  10.1103/PhysRevLett.106.057403} {\bibfield  {journal} {\bibinfo  {journal}
  {Phys. Rev. Lett.}\ }\textbf {\bibinfo {volume} {106}},\ \bibinfo {pages}
  {057403} (\bibinfo {year} {2011})}\BibitemShut {NoStop}%
\bibitem [{\citenamefont {Bordacs}\ \emph {et~al.}(2012)\citenamefont
  {Bordacs}, \citenamefont {Kezsmarki}, \citenamefont {Szaller}, \citenamefont
  {Demko}, \citenamefont {Kida}, \citenamefont {Murakawa}, \citenamefont
  {Onose}, \citenamefont {Shimano}, \citenamefont {R{\~o}{\~o}m}, \citenamefont
  {Nagel}, \citenamefont {Miyahara}, \citenamefont {Furukawa},\ and\
  \citenamefont {Tokura}}]{bordacs_nphys_2012}%
  \BibitemOpen
  \bibfield  {author} {\bibinfo {author} {\bibfnamefont {S.}~\bibnamefont
  {Bordacs}}, \bibinfo {author} {\bibfnamefont {I.}~\bibnamefont {Kezsmarki}},
  \bibinfo {author} {\bibfnamefont {D.}~\bibnamefont {Szaller}}, \bibinfo
  {author} {\bibfnamefont {L.}~\bibnamefont {Demko}}, \bibinfo {author}
  {\bibfnamefont {N.}~\bibnamefont {Kida}}, \bibinfo {author} {\bibfnamefont
  {H.}~\bibnamefont {Murakawa}}, \bibinfo {author} {\bibfnamefont
  {Y.}~\bibnamefont {Onose}}, \bibinfo {author} {\bibfnamefont
  {R.}~\bibnamefont {Shimano}}, \bibinfo {author} {\bibfnamefont
  {T.}~\bibnamefont {R{\~o}{\~o}m}}, \bibinfo {author} {\bibfnamefont
  {U.}~\bibnamefont {Nagel}}, \bibinfo {author} {\bibfnamefont
  {S.}~\bibnamefont {Miyahara}}, \bibinfo {author} {\bibfnamefont
  {N.}~\bibnamefont {Furukawa}}, \ and\ \bibinfo {author} {\bibfnamefont
  {Y.}~\bibnamefont {Tokura}},\ }\bibfield  {title} {\enquote {\bibinfo {title}
  {Chirality of matter shows up via spin excitations},}\ }\href {\doibase
  10.1038/NPHYS2387} {\bibfield  {journal} {\bibinfo  {journal} {Nat. Phys.}\
  }\textbf {\bibinfo {volume} {8}},\ \bibinfo {pages} {734} (\bibinfo {year}
  {2012})}\BibitemShut {NoStop}%
\bibitem [{\citenamefont {Takahashi}\ \emph {et~al.}(2012)\citenamefont
  {Takahashi}, \citenamefont {Shimano}, \citenamefont {Kaneko}, \citenamefont
  {Murakawa},\ and\ \citenamefont {Tokura}}]{takahashi_nphys_2012}%
  \BibitemOpen
  \bibfield  {author} {\bibinfo {author} {\bibfnamefont {Y.}~\bibnamefont
  {Takahashi}}, \bibinfo {author} {\bibfnamefont {R.}~\bibnamefont {Shimano}},
  \bibinfo {author} {\bibfnamefont {Y.}~\bibnamefont {Kaneko}}, \bibinfo
  {author} {\bibfnamefont {H.}~\bibnamefont {Murakawa}}, \ and\ \bibinfo
  {author} {\bibfnamefont {Y.}~\bibnamefont {Tokura}},\ }\bibfield  {title}
  {\enquote {\bibinfo {title} {Magnetoelectric resonance with electromagnons in
  a perovskite helimagnet},}\ }\href {\doibase 10.1038/NPHYS2161} {\bibfield
  {journal} {\bibinfo  {journal} {Nat. Phys.}\ }\textbf {\bibinfo {volume}
  {8}},\ \bibinfo {pages} {121} (\bibinfo {year} {2012})}\BibitemShut {NoStop}%
\bibitem [{\citenamefont {Takahashi}\ \emph {et~al.}(2013)\citenamefont
  {Takahashi}, \citenamefont {Yamasaki},\ and\ \citenamefont
  {Tokura}}]{takahashi_prl_2013}%
  \BibitemOpen
  \bibfield  {author} {\bibinfo {author} {\bibfnamefont {Y.}~\bibnamefont
  {Takahashi}}, \bibinfo {author} {\bibfnamefont {Y.}~\bibnamefont {Yamasaki}},
  \ and\ \bibinfo {author} {\bibfnamefont {Y.}~\bibnamefont {Tokura}},\
  }\bibfield  {title} {\enquote {\bibinfo {title} {Terahertz magnetoelectric
  resonance enhanced by mutual coupling of electromagnons},}\ }\href {\doibase
  10.1103/PhysRevLett.111.037204} {\bibfield  {journal} {\bibinfo  {journal}
  {Phys. Rev. Lett.}\ }\textbf {\bibinfo {volume} {111}},\ \bibinfo {pages}
  {037204} (\bibinfo {year} {2013})}\BibitemShut {NoStop}%
\bibitem [{\citenamefont {Szaller}\ \emph {et~al.}(2013)\citenamefont
  {Szaller}, \citenamefont {Bord\'acs},\ and\ \citenamefont
  {K\'ezsm\'arki}}]{szaller_prb_2013}%
  \BibitemOpen
  \bibfield  {author} {\bibinfo {author} {\bibfnamefont {D.}~\bibnamefont
  {Szaller}}, \bibinfo {author} {\bibfnamefont {S.}~\bibnamefont {Bord\'acs}},
  \ and\ \bibinfo {author} {\bibfnamefont {I.}~\bibnamefont {K\'ezsm\'arki}},\
  }\bibfield  {title} {\enquote {\bibinfo {title} {Symmetry conditions for
  nonreciprocal light propagation in magnetic crystals},}\ }\href {\doibase
  10.1103/PhysRevB.87.014421} {\bibfield  {journal} {\bibinfo  {journal} {Phys.
  Rev. B}\ }\textbf {\bibinfo {volume} {87}},\ \bibinfo {pages} {014421}
  (\bibinfo {year} {2013})}\BibitemShut {NoStop}%
\bibitem [{\citenamefont {K\'ezsm\'arki}\ \emph {et~al.}(2014)\citenamefont
  {K\'ezsm\'arki}, \citenamefont {Szaller}, \citenamefont {Bord\'acs},
  \citenamefont {Kocsis}, \citenamefont {Tokunaga}, \citenamefont {Taguchi},
  \citenamefont {Murakawa}, \citenamefont {Tokura}, \citenamefont {Engelkamp},
  \citenamefont {R{\~o}{\~o}m},\ and\ \citenamefont
  {Nagel}}]{kezsmarki_nc_2014}%
  \BibitemOpen
  \bibfield  {author} {\bibinfo {author} {\bibfnamefont {I.}~\bibnamefont
  {K\'ezsm\'arki}}, \bibinfo {author} {\bibfnamefont {D.}~\bibnamefont
  {Szaller}}, \bibinfo {author} {\bibfnamefont {S.}~\bibnamefont {Bord\'acs}},
  \bibinfo {author} {\bibfnamefont {V.}~\bibnamefont {Kocsis}}, \bibinfo
  {author} {\bibfnamefont {Y.}~\bibnamefont {Tokunaga}}, \bibinfo {author}
  {\bibfnamefont {Y.}~\bibnamefont {Taguchi}}, \bibinfo {author} {\bibfnamefont
  {H.}~\bibnamefont {Murakawa}}, \bibinfo {author} {\bibfnamefont
  {Y.}~\bibnamefont {Tokura}}, \bibinfo {author} {\bibfnamefont
  {H.}~\bibnamefont {Engelkamp}}, \bibinfo {author} {\bibfnamefont
  {T.}~\bibnamefont {R{\~o}{\~o}m}}, \ and\ \bibinfo {author} {\bibfnamefont
  {U.}~\bibnamefont {Nagel}},\ }\bibfield  {title} {\enquote {\bibinfo {title}
  {One-way transparency of four-coloured spin-wave excitations in multiferroic
  materials},}\ }\href {\doibase 10.1038/ncomms4203} {\bibfield  {journal}
  {\bibinfo  {journal} {Nat. Commun.}\ }\textbf {\bibinfo {volume} {5}},\
  \bibinfo {pages} {3203} (\bibinfo {year} {2014})}\BibitemShut {NoStop}%
\bibitem [{\citenamefont {Szaller}\ \emph {et~al.}(2014)\citenamefont
  {Szaller}, \citenamefont {Bord\'acs}, \citenamefont {Kocsis}, \citenamefont
  {R{\~o}{\~o}m}, \citenamefont {Nagel},\ and\ \citenamefont
  {K\'ezsm\'arki}}]{szaller_prb_2014}%
  \BibitemOpen
  \bibfield  {author} {\bibinfo {author} {\bibfnamefont {D.}~\bibnamefont
  {Szaller}}, \bibinfo {author} {\bibfnamefont {S.}~\bibnamefont {Bord\'acs}},
  \bibinfo {author} {\bibfnamefont {V.}~\bibnamefont {Kocsis}}, \bibinfo
  {author} {\bibfnamefont {T.}~\bibnamefont {R{\~o}{\~o}m}}, \bibinfo {author}
  {\bibfnamefont {U.}~\bibnamefont {Nagel}}, \ and\ \bibinfo {author}
  {\bibfnamefont {I.}~\bibnamefont {K\'ezsm\'arki}},\ }\bibfield  {title}
  {\enquote {\bibinfo {title} {Effect of spin excitations with simultaneous
  magnetic- and electric-dipole character on the static magnetoelectric
  properties of multiferroic materials},}\ }\href {\doibase
  10.1103/PhysRevB.89.184419} {\bibfield  {journal} {\bibinfo  {journal} {Phys.
  Rev. B}\ }\textbf {\bibinfo {volume} {89}},\ \bibinfo {pages} {184419}
  (\bibinfo {year} {2014})}\BibitemShut {NoStop}%
\bibitem [{\citenamefont {Kuzmenko}\ \emph {et~al.}(2015)\citenamefont
  {Kuzmenko}, \citenamefont {Dziom}, \citenamefont {Shuvaev}, \citenamefont
  {Pimenov}, \citenamefont {Schiebl}, \citenamefont {Mukhin}, \citenamefont
  {Ivanov}, \citenamefont {Gudim}, \citenamefont {Bezmaternykh},\ and\
  \citenamefont {Pimenov}}]{kuzmenko_prb_2015}%
  \BibitemOpen
  \bibfield  {author} {\bibinfo {author} {\bibfnamefont {A.~M.}\ \bibnamefont
  {Kuzmenko}}, \bibinfo {author} {\bibfnamefont {V.}~\bibnamefont {Dziom}},
  \bibinfo {author} {\bibfnamefont {A.}~\bibnamefont {Shuvaev}}, \bibinfo
  {author} {\bibfnamefont {Anna}\ \bibnamefont {Pimenov}}, \bibinfo {author}
  {\bibfnamefont {M.}~\bibnamefont {Schiebl}}, \bibinfo {author} {\bibfnamefont
  {A.~A.}\ \bibnamefont {Mukhin}}, \bibinfo {author} {\bibfnamefont {V.~Yu.}\
  \bibnamefont {Ivanov}}, \bibinfo {author} {\bibfnamefont {I.~A.}\
  \bibnamefont {Gudim}}, \bibinfo {author} {\bibfnamefont {L.~N.}\ \bibnamefont
  {Bezmaternykh}}, \ and\ \bibinfo {author} {\bibfnamefont {A.}~\bibnamefont
  {Pimenov}},\ }\bibfield  {title} {\enquote {\bibinfo {title} {Large
  directional optical anisotropy in multiferroic ferroborate},}\ }\href
  {\doibase 10.1103/PhysRevB.92.184409} {\bibfield  {journal} {\bibinfo
  {journal} {Phys. Rev. B}\ }\textbf {\bibinfo {volume} {92}},\ \bibinfo
  {pages} {184409} (\bibinfo {year} {2015})}\BibitemShut {NoStop}%
\bibitem [{\citenamefont {K\'ezsm\'arki}\ \emph {et~al.}(2015)\citenamefont
  {K\'ezsm\'arki}, \citenamefont {Nagel}, \citenamefont {Bord\'acs},
  \citenamefont {Fishman}, \citenamefont {Lee}, \citenamefont {Yi},
  \citenamefont {Cheong},\ and\ \citenamefont
  {R{\~o}{\~o}m}}]{kezsmarki_prl_2015}%
  \BibitemOpen
  \bibfield  {author} {\bibinfo {author} {\bibfnamefont {I.}~\bibnamefont
  {K\'ezsm\'arki}}, \bibinfo {author} {\bibfnamefont {U.}~\bibnamefont
  {Nagel}}, \bibinfo {author} {\bibfnamefont {S.}~\bibnamefont {Bord\'acs}},
  \bibinfo {author} {\bibfnamefont {R.~S.}\ \bibnamefont {Fishman}}, \bibinfo
  {author} {\bibfnamefont {J.~H.}\ \bibnamefont {Lee}}, \bibinfo {author}
  {\bibfnamefont {H.~T.}\ \bibnamefont {Yi}}, \bibinfo {author} {\bibfnamefont
  {S.-W.}\ \bibnamefont {Cheong}}, \ and\ \bibinfo {author} {\bibfnamefont
  {T.}~\bibnamefont {R{\~o}{\~o}m}},\ }\bibfield  {title} {\enquote {\bibinfo
  {title} {Optical diode effect at spin-wave excitations of the
  room-temperature multiferroic {BiFeO}$_3$},}\ }\href {\doibase
  10.1103/PhysRevLett.115.127203} {\bibfield  {journal} {\bibinfo  {journal}
  {Phys. Rev. Lett.}\ }\textbf {\bibinfo {volume} {115}},\ \bibinfo {pages}
  {127203} (\bibinfo {year} {2015})}\BibitemShut {NoStop}%
\bibitem [{\citenamefont {Kuzmenko}\ \emph {et~al.}(2014)\citenamefont
  {Kuzmenko}, \citenamefont {Shuvaev}, \citenamefont {Dziom}, \citenamefont
  {Pimenov}, \citenamefont {Schiebl}, \citenamefont {Mukhin}, \citenamefont
  {Ivanov}, \citenamefont {Bezmaternykh},\ and\ \citenamefont
  {Pimenov}}]{kuzmenko_prb_2014}%
  \BibitemOpen
  \bibfield  {author} {\bibinfo {author} {\bibfnamefont {A.~M.}\ \bibnamefont
  {Kuzmenko}}, \bibinfo {author} {\bibfnamefont {A.}~\bibnamefont {Shuvaev}},
  \bibinfo {author} {\bibfnamefont {V.}~\bibnamefont {Dziom}}, \bibinfo
  {author} {\bibfnamefont {Anna}\ \bibnamefont {Pimenov}}, \bibinfo {author}
  {\bibfnamefont {M.}~\bibnamefont {Schiebl}}, \bibinfo {author} {\bibfnamefont
  {A.~A.}\ \bibnamefont {Mukhin}}, \bibinfo {author} {\bibfnamefont {V.~Yu.}\
  \bibnamefont {Ivanov}}, \bibinfo {author} {\bibfnamefont {L.~N.}\
  \bibnamefont {Bezmaternykh}}, \ and\ \bibinfo {author} {\bibfnamefont
  {A.}~\bibnamefont {Pimenov}},\ }\bibfield  {title} {\enquote {\bibinfo
  {title} {Giant gigahertz optical activity in multiferroic ferroborate},}\
  }\href {\doibase 10.1103/PhysRevB.89.174407} {\bibfield  {journal} {\bibinfo
  {journal} {Phys. Rev. B}\ }\textbf {\bibinfo {volume} {89}},\ \bibinfo
  {pages} {174407} (\bibinfo {year} {2014})}\BibitemShut {NoStop}%
\bibitem [{\citenamefont {Kurumaji}\ \emph
  {et~al.}(2017{\natexlab{a}})\citenamefont {Kurumaji}, \citenamefont
  {Takahashi}, \citenamefont {Fujioka}, \citenamefont {Masuda}, \citenamefont
  {Shishikura}, \citenamefont {Ishiwata},\ and\ \citenamefont
  {Tokura}}]{kurumaji_prl_2017}%
  \BibitemOpen
  \bibfield  {author} {\bibinfo {author} {\bibfnamefont {T.}~\bibnamefont
  {Kurumaji}}, \bibinfo {author} {\bibfnamefont {Y.}~\bibnamefont {Takahashi}},
  \bibinfo {author} {\bibfnamefont {J.}~\bibnamefont {Fujioka}}, \bibinfo
  {author} {\bibfnamefont {R.}~\bibnamefont {Masuda}}, \bibinfo {author}
  {\bibfnamefont {H.}~\bibnamefont {Shishikura}}, \bibinfo {author}
  {\bibfnamefont {S.}~\bibnamefont {Ishiwata}}, \ and\ \bibinfo {author}
  {\bibfnamefont {Y.}~\bibnamefont {Tokura}},\ }\bibfield  {title} {\enquote
  {\bibinfo {title} {Optical magnetoelectric resonance in a polar magnet
  ({Fe},{Zn})$_{2}${Mo}$_{3}${O}$_{8}$ with axion-type coupling},}\ }\href
  {\doibase 10.1103/PhysRevLett.119.077206} {\bibfield  {journal} {\bibinfo
  {journal} {Phys. Rev. Lett.}\ }\textbf {\bibinfo {volume} {119}},\ \bibinfo
  {pages} {077206} (\bibinfo {year} {2017}{\natexlab{a}})}\BibitemShut
  {NoStop}%
\bibitem [{\citenamefont {Yu}\ \emph {et~al.}(2018)\citenamefont {Yu},
  \citenamefont {Gao}, \citenamefont {Kim}, \citenamefont {Cheong},
  \citenamefont {Man}, \citenamefont {Mad\'eo}, \citenamefont {Dani},\ and\
  \citenamefont {Talbayev}}]{yu_prl_2018}%
  \BibitemOpen
  \bibfield  {author} {\bibinfo {author} {\bibfnamefont {Shukai}\ \bibnamefont
  {Yu}}, \bibinfo {author} {\bibfnamefont {Bin}\ \bibnamefont {Gao}}, \bibinfo
  {author} {\bibfnamefont {Jae~Wook}\ \bibnamefont {Kim}}, \bibinfo {author}
  {\bibfnamefont {Sang-Wook}\ \bibnamefont {Cheong}}, \bibinfo {author}
  {\bibfnamefont {Michael K.~L.}\ \bibnamefont {Man}}, \bibinfo {author}
  {\bibfnamefont {Julien}\ \bibnamefont {Mad\'eo}}, \bibinfo {author}
  {\bibfnamefont {Keshav~M.}\ \bibnamefont {Dani}}, \ and\ \bibinfo {author}
  {\bibfnamefont {Diyar}\ \bibnamefont {Talbayev}},\ }\bibfield  {title}
  {\enquote {\bibinfo {title} {High-temperature terahertz optical diode effect
  without magnetic order in polar {FeZnMo$_3$O$_8$}},}\ }\href {\doibase
  10.1103/PhysRevLett.120.037601} {\bibfield  {journal} {\bibinfo  {journal}
  {Phys. Rev. Lett.}\ }\textbf {\bibinfo {volume} {120}},\ \bibinfo {pages}
  {037601} (\bibinfo {year} {2018})}\BibitemShut {NoStop}%
\bibitem [{\citenamefont {Viirok}\ \emph {et~al.}(2019)\citenamefont {Viirok},
  \citenamefont {Nagel}, \citenamefont {R{\~o}{\~o}m}, \citenamefont {Farkas},
  \citenamefont {Balla}, \citenamefont {Szaller}, \citenamefont {Kocsis},
  \citenamefont {Tokunaga}, \citenamefont {Taguchi}, \citenamefont {Tokura},
  \citenamefont {Bern\'ath}, \citenamefont {Kamenskyi}, \citenamefont
  {K\'ezsm\'arki}, \citenamefont {Bord\'acs},\ and\ \citenamefont
  {Penc}}]{viirok_prb_2019}%
  \BibitemOpen
  \bibfield  {author} {\bibinfo {author} {\bibfnamefont {J.}~\bibnamefont
  {Viirok}}, \bibinfo {author} {\bibfnamefont {U.}~\bibnamefont {Nagel}},
  \bibinfo {author} {\bibfnamefont {T.}~\bibnamefont {R{\~o}{\~o}m}}, \bibinfo
  {author} {\bibfnamefont {D.~G.}\ \bibnamefont {Farkas}}, \bibinfo {author}
  {\bibfnamefont {P.}~\bibnamefont {Balla}}, \bibinfo {author} {\bibfnamefont
  {D.}~\bibnamefont {Szaller}}, \bibinfo {author} {\bibfnamefont
  {V.}~\bibnamefont {Kocsis}}, \bibinfo {author} {\bibfnamefont
  {Y.}~\bibnamefont {Tokunaga}}, \bibinfo {author} {\bibfnamefont
  {Y.}~\bibnamefont {Taguchi}}, \bibinfo {author} {\bibfnamefont
  {Y.}~\bibnamefont {Tokura}}, \bibinfo {author} {\bibfnamefont
  {B.}~\bibnamefont {Bern\'ath}}, \bibinfo {author} {\bibfnamefont {D.~L.}\
  \bibnamefont {Kamenskyi}}, \bibinfo {author} {\bibfnamefont {I.}~\bibnamefont
  {K\'ezsm\'arki}}, \bibinfo {author} {\bibfnamefont {S.}~\bibnamefont
  {Bord\'acs}}, \ and\ \bibinfo {author} {\bibfnamefont {K.}~\bibnamefont
  {Penc}},\ }\bibfield  {title} {\enquote {\bibinfo {title} {Directional
  dichroism in the paramagnetic state of multiferroics: A case study of
  infrared light absorption in {Sr$_2$CoSi$_2$O$_7$} at high temperatures},}\
  }\href {\doibase 10.1103/PhysRevB.99.014410} {\bibfield  {journal} {\bibinfo
  {journal} {Phys. Rev. B}\ }\textbf {\bibinfo {volume} {99}},\ \bibinfo
  {pages} {014410} (\bibinfo {year} {2019})}\BibitemShut {NoStop}%
\bibitem [{\citenamefont {Kuzmenko}\ \emph {et~al.}(2019)\citenamefont
  {Kuzmenko}, \citenamefont {Dziom}, \citenamefont {Shuvaev}, \citenamefont
  {Pimenov}, \citenamefont {Szaller}, \citenamefont {Mukhin}, \citenamefont
  {Ivanov},\ and\ \citenamefont {Pimenov}}]{kuzmenko_prb_2019}%
  \BibitemOpen
  \bibfield  {author} {\bibinfo {author} {\bibfnamefont {A.~M.}\ \bibnamefont
  {Kuzmenko}}, \bibinfo {author} {\bibfnamefont {V.}~\bibnamefont {Dziom}},
  \bibinfo {author} {\bibfnamefont {A.}~\bibnamefont {Shuvaev}}, \bibinfo
  {author} {\bibfnamefont {Anna}\ \bibnamefont {Pimenov}}, \bibinfo {author}
  {\bibfnamefont {D.}~\bibnamefont {Szaller}}, \bibinfo {author} {\bibfnamefont
  {A.~A.}\ \bibnamefont {Mukhin}}, \bibinfo {author} {\bibfnamefont {V.~Yu.}\
  \bibnamefont {Ivanov}}, \ and\ \bibinfo {author} {\bibfnamefont
  {A.}~\bibnamefont {Pimenov}},\ }\bibfield  {title} {\enquote {\bibinfo
  {title} {Sign change of polarization rotation under time or space inversion
  in magnetoelectric {YbAl$_3$(BO$_3)_4$}},}\ }\href {\doibase
  10.1103/PhysRevB.99.224417} {\bibfield  {journal} {\bibinfo  {journal} {Phys.
  Rev. B}\ }\textbf {\bibinfo {volume} {99}},\ \bibinfo {pages} {224417}
  (\bibinfo {year} {2019})}\BibitemShut {NoStop}%
\bibitem [{\citenamefont {Yokosuk}\ \emph {et~al.}(2020)\citenamefont
  {Yokosuk}, \citenamefont {Kim}, \citenamefont {Hughey}, \citenamefont {Kim},
  \citenamefont {Stier}, \citenamefont {O’Neal}, \citenamefont {Yang},
  \citenamefont {Crooker}, \citenamefont {Haule}, \citenamefont {Cheong} \emph
  {et~al.}}]{yokosuk_npjqm_2020}%
  \BibitemOpen
  \bibfield  {author} {\bibinfo {author} {\bibfnamefont {Michael~O}\
  \bibnamefont {Yokosuk}}, \bibinfo {author} {\bibfnamefont {Heung-Sik}\
  \bibnamefont {Kim}}, \bibinfo {author} {\bibfnamefont {Kendall~D}\
  \bibnamefont {Hughey}}, \bibinfo {author} {\bibfnamefont {Jaewook}\
  \bibnamefont {Kim}}, \bibinfo {author} {\bibfnamefont {Andreas~V}\
  \bibnamefont {Stier}}, \bibinfo {author} {\bibfnamefont {Kenneth~R}\
  \bibnamefont {O’Neal}}, \bibinfo {author} {\bibfnamefont {Junjie}\
  \bibnamefont {Yang}}, \bibinfo {author} {\bibfnamefont {Scott~A}\
  \bibnamefont {Crooker}}, \bibinfo {author} {\bibfnamefont {Kristjan}\
  \bibnamefont {Haule}}, \bibinfo {author} {\bibfnamefont {Sang-Wook}\
  \bibnamefont {Cheong}},  \emph {et~al.},\ }\bibfield  {title} {\enquote
  {\bibinfo {title} {Nonreciprocal directional dichroism of a chiral magnet in
  the visible range},}\ }\href@noop {} {\bibfield  {journal} {\bibinfo
  {journal} {npj Quantum Materials}\ }\textbf {\bibinfo {volume} {5}},\
  \bibinfo {pages} {1--8} (\bibinfo {year} {2020})}\BibitemShut {NoStop}%
\bibitem [{\citenamefont {Katsura}\ \emph {et~al.}(2005)\citenamefont
  {Katsura}, \citenamefont {Nagaosa},\ and\ \citenamefont
  {Balatsky}}]{katsura_prl_2005}%
  \BibitemOpen
  \bibfield  {author} {\bibinfo {author} {\bibfnamefont {H.}~\bibnamefont
  {Katsura}}, \bibinfo {author} {\bibfnamefont {N.}~\bibnamefont {Nagaosa}}, \
  and\ \bibinfo {author} {\bibfnamefont {A.~V.}\ \bibnamefont {Balatsky}},\
  }\bibfield  {title} {\enquote {\bibinfo {title} {Spin current and
  magnetoelectric effect in noncollinear magnets},}\ }\href {\doibase
  10.1103/PhysRevLett.95.057205} {\bibfield  {journal} {\bibinfo  {journal}
  {Phys. Rev. Lett.}\ }\textbf {\bibinfo {volume} {95}},\ \bibinfo {pages}
  {057205} (\bibinfo {year} {2005})}\BibitemShut {NoStop}%
\bibitem [{\citenamefont {Jia}\ \emph {et~al.}(2007)\citenamefont {Jia},
  \citenamefont {Onoda}, \citenamefont {Nagaosa},\ and\ \citenamefont
  {Han}}]{jia_prb_2007}%
  \BibitemOpen
  \bibfield  {author} {\bibinfo {author} {\bibfnamefont {Chenglong}\
  \bibnamefont {Jia}}, \bibinfo {author} {\bibfnamefont {Shigeki}\ \bibnamefont
  {Onoda}}, \bibinfo {author} {\bibfnamefont {Naoto}\ \bibnamefont {Nagaosa}},
  \ and\ \bibinfo {author} {\bibfnamefont {Jung~Hoon}\ \bibnamefont {Han}},\
  }\bibfield  {title} {\enquote {\bibinfo {title} {Microscopic theory of
  spin-polarization coupling in multiferroic transition metal oxides},}\ }\href
  {\doibase 10.1103/PhysRevB.76.144424} {\bibfield  {journal} {\bibinfo
  {journal} {Phys. Rev. B}\ }\textbf {\bibinfo {volume} {76}},\ \bibinfo
  {pages} {144424} (\bibinfo {year} {2007})}\BibitemShut {NoStop}%
\bibitem [{\citenamefont {Arima}(2008)}]{arima_jpcm_2008}%
  \BibitemOpen
  \bibfield  {author} {\bibinfo {author} {\bibfnamefont {T}~\bibnamefont
  {Arima}},\ }\bibfield  {title} {\enquote {\bibinfo {title} {Magneto-electric
  optics in non-centrosymmetric ferromagnets},}\ }\href@noop {} {\bibfield
  {journal} {\bibinfo  {journal} {J. Phys.: Condens. Matter}\ }\textbf
  {\bibinfo {volume} {20}},\ \bibinfo {pages} {434211} (\bibinfo {year}
  {2008})}\BibitemShut {NoStop}%
\bibitem [{\citenamefont {Murakawa}\ \emph {et~al.}(2010)\citenamefont
  {Murakawa}, \citenamefont {Onose}, \citenamefont {Miyahara}, \citenamefont
  {Furukawa},\ and\ \citenamefont {Tokura}}]{murakawa_prl_2010}%
  \BibitemOpen
  \bibfield  {author} {\bibinfo {author} {\bibfnamefont {H.}~\bibnamefont
  {Murakawa}}, \bibinfo {author} {\bibfnamefont {Y.}~\bibnamefont {Onose}},
  \bibinfo {author} {\bibfnamefont {S.}~\bibnamefont {Miyahara}}, \bibinfo
  {author} {\bibfnamefont {N.}~\bibnamefont {Furukawa}}, \ and\ \bibinfo
  {author} {\bibfnamefont {Y.}~\bibnamefont {Tokura}},\ }\bibfield  {title}
  {\enquote {\bibinfo {title} {Ferroelectricity induced by spin-dependent
  metal-ligand hybridization in {Ba}$_{2}${CoGe}$_{2}${O}$_{7}$},}\ }\href
  {\doibase 10.1103/PhysRevLett.105.137202} {\bibfield  {journal} {\bibinfo
  {journal} {Phys. Rev. Lett.}\ }\textbf {\bibinfo {volume} {105}},\ \bibinfo
  {pages} {137202} (\bibinfo {year} {2010})}\BibitemShut {NoStop}%
\bibitem [{\citenamefont {Sergienko}\ \emph {et~al.}(2006)\citenamefont
  {Sergienko}, \citenamefont {Sen},\ and\ \citenamefont
  {Dagotto}}]{sergienko_prl_2006}%
  \BibitemOpen
  \bibfield  {author} {\bibinfo {author} {\bibfnamefont {Ivan~A.}\ \bibnamefont
  {Sergienko}}, \bibinfo {author} {\bibfnamefont {Cengiz}\ \bibnamefont {Sen}},
  \ and\ \bibinfo {author} {\bibfnamefont {Elbio}\ \bibnamefont {Dagotto}},\
  }\bibfield  {title} {\enquote {\bibinfo {title} {Ferroelectricity in the
  magnetic e-phase of orthorhombic perovskites},}\ }\href {\doibase
  10.1103/PhysRevLett.97.227204} {\bibfield  {journal} {\bibinfo  {journal}
  {Phys. Rev. Lett.}\ }\textbf {\bibinfo {volume} {97}},\ \bibinfo {eid}
  {227204} (\bibinfo {year} {2006})}\BibitemShut {NoStop}%
\bibitem [{\citenamefont {Choi}\ \emph {et~al.}(2008)\citenamefont {Choi},
  \citenamefont {Yi}, \citenamefont {Lee}, \citenamefont {Huang}, \citenamefont
  {Kiryukhin},\ and\ \citenamefont {Cheong}}]{choi_prl_2008}%
  \BibitemOpen
  \bibfield  {author} {\bibinfo {author} {\bibfnamefont {Y.~J.}\ \bibnamefont
  {Choi}}, \bibinfo {author} {\bibfnamefont {H.~T.}\ \bibnamefont {Yi}},
  \bibinfo {author} {\bibfnamefont {S.}~\bibnamefont {Lee}}, \bibinfo {author}
  {\bibfnamefont {Q.}~\bibnamefont {Huang}}, \bibinfo {author} {\bibfnamefont
  {V.}~\bibnamefont {Kiryukhin}}, \ and\ \bibinfo {author} {\bibfnamefont
  {S.-W.}\ \bibnamefont {Cheong}},\ }\bibfield  {title} {\enquote {\bibinfo
  {title} {Ferroelectricity in an {Ising} chain magnet},}\ }\href {\doibase
  10.1103/PhysRevLett.100.047601} {\bibfield  {journal} {\bibinfo  {journal}
  {Phys. Rev. Lett.}\ }\textbf {\bibinfo {volume} {100}},\ \bibinfo {pages}
  {047601} (\bibinfo {year} {2008})}\BibitemShut {NoStop}%
\bibitem [{\citenamefont {Wang}\ \emph {et~al.}(2015)\citenamefont {Wang},
  \citenamefont {Pascut}, \citenamefont {Gao}, \citenamefont {Tyson},
  \citenamefont {Haule}, \citenamefont {Kiryukhin},\ and\ \citenamefont
  {Cheong}}]{wang_scirep_2015}%
  \BibitemOpen
  \bibfield  {author} {\bibinfo {author} {\bibfnamefont {Yazhong}\ \bibnamefont
  {Wang}}, \bibinfo {author} {\bibfnamefont {Gheorghe~L}\ \bibnamefont
  {Pascut}}, \bibinfo {author} {\bibfnamefont {Bin}\ \bibnamefont {Gao}},
  \bibinfo {author} {\bibfnamefont {Trevor~A}\ \bibnamefont {Tyson}}, \bibinfo
  {author} {\bibfnamefont {Kristjan}\ \bibnamefont {Haule}}, \bibinfo {author}
  {\bibfnamefont {Valery}\ \bibnamefont {Kiryukhin}}, \ and\ \bibinfo {author}
  {\bibfnamefont {Sang-Wook}\ \bibnamefont {Cheong}},\ }\bibfield  {title}
  {\enquote {\bibinfo {title} {Unveiling hidden ferrimagnetism and giant
  magnetoelectricity in polar magnet {Fe$_2$Mo$_3$O$_8$}},}\ }\href@noop {}
  {\bibfield  {journal} {\bibinfo  {journal} {Scientific reports}\ }\textbf
  {\bibinfo {volume} {5}},\ \bibinfo {pages} {12268} (\bibinfo {year}
  {2015})}\BibitemShut {NoStop}%
\bibitem [{\citenamefont {Khomskii}(2006)}]{khomskii_jmmm_2006}%
  \BibitemOpen
  \bibfield  {author} {\bibinfo {author} {\bibfnamefont {D.~I.}\ \bibnamefont
  {Khomskii}},\ }\bibfield  {title} {\enquote {\bibinfo {title} {Multiferroics:
  Different ways to combine magnetism and ferroelectricity},}\ }\href {\doibase
  DOI: 10.1016/j.jmmm.2006.01.238} {\bibfield  {journal} {\bibinfo  {journal}
  {J. Magn. Magn. Mater.}\ }\textbf {\bibinfo {volume} {306}},\ \bibinfo
  {pages} {1} (\bibinfo {year} {2006})}\BibitemShut {NoStop}%
\bibitem [{\citenamefont {Cotton}(1966)}]{cotton_qrcs_1966}%
  \BibitemOpen
  \bibfield  {author} {\bibinfo {author} {\bibfnamefont {FA}~\bibnamefont
  {Cotton}},\ }\bibfield  {title} {\enquote {\bibinfo {title} {Transition-metal
  compounds containing clusters of metal atoms},}\ }\href@noop {} {\bibfield
  {journal} {\bibinfo  {journal} {Quarterly Reviews, Chemical Society}\
  }\textbf {\bibinfo {volume} {20}},\ \bibinfo {pages} {389--401} (\bibinfo
  {year} {1966})}\BibitemShut {NoStop}%
\bibitem [{\citenamefont {Varret}\ \emph {et~al.}(1972)\citenamefont {Varret},
  \citenamefont {Czeskleba}, \citenamefont {Hartmann-Boutron},\ and\
  \citenamefont {Imbert}}]{varret_jp_1972}%
  \BibitemOpen
  \bibfield  {author} {\bibinfo {author} {\bibfnamefont {F}~\bibnamefont
  {Varret}}, \bibinfo {author} {\bibfnamefont {H}~\bibnamefont {Czeskleba}},
  \bibinfo {author} {\bibfnamefont {F}~\bibnamefont {Hartmann-Boutron}}, \ and\
  \bibinfo {author} {\bibfnamefont {P}~\bibnamefont {Imbert}},\ }\bibfield
  {title} {\enquote {\bibinfo {title} {{\'E}tude par effet {M{\"o}ssbauer} de
  l'ion {Fe$^{2+}$} en sym{\'e}trie trigonale dans les compos{\'e}s du type
  {(Fe,M)$_2$Mo$_3$O$_8$ (M = Mg, Zn, Mn, Co, Ni)} et propri{\'e}t{\'e}s
  magn{\'e}tiques de {(Fe, Zn)$_2$Mo$_3$O$_8$}},}\ }\href@noop {} {\bibfield
  {journal} {\bibinfo  {journal} {Journal de Physique}\ }\textbf {\bibinfo
  {volume} {33}},\ \bibinfo {pages} {549--564} (\bibinfo {year}
  {1972})}\BibitemShut {NoStop}%
\bibitem [{\citenamefont {McAlister}\ and\ \citenamefont
  {Strobel}(1983)}]{mcalister_jmmm_1983}%
  \BibitemOpen
  \bibfield  {author} {\bibinfo {author} {\bibfnamefont {SP}~\bibnamefont
  {McAlister}}\ and\ \bibinfo {author} {\bibfnamefont {P}~\bibnamefont
  {Strobel}},\ }\bibfield  {title} {\enquote {\bibinfo {title} {Magnetic order
  in {M$_2$Mo$_3$O$_8$} single crystals {(M = Mn, Fe, Co, Ni)}},}\ }\href@noop
  {} {\bibfield  {journal} {\bibinfo  {journal} {Journal of Magnetism and
  Magnetic Materials}\ }\textbf {\bibinfo {volume} {30}},\ \bibinfo {pages}
  {340--348} (\bibinfo {year} {1983})}\BibitemShut {NoStop}%
\bibitem [{\citenamefont {Kurumaji}\ \emph {et~al.}(2015)\citenamefont
  {Kurumaji}, \citenamefont {Ishiwata},\ and\ \citenamefont
  {Tokura}}]{kurumaji_prx_2015}%
  \BibitemOpen
  \bibfield  {author} {\bibinfo {author} {\bibfnamefont {T.}~\bibnamefont
  {Kurumaji}}, \bibinfo {author} {\bibfnamefont {S.}~\bibnamefont {Ishiwata}},
  \ and\ \bibinfo {author} {\bibfnamefont {Y.}~\bibnamefont {Tokura}},\
  }\bibfield  {title} {\enquote {\bibinfo {title} {Doping-tunable ferrimagnetic
  phase with large linear magnetoelectric effect in a polar magnet
  ${\mathrm{fe}}_{2}{\mathrm{mo}}_{3}{\mathrm{o}}_{8}$},}\ }\href {\doibase
  10.1103/PhysRevX.5.031034} {\bibfield  {journal} {\bibinfo  {journal} {Phys.
  Rev. X}\ }\textbf {\bibinfo {volume} {5}},\ \bibinfo {pages} {031034}
  (\bibinfo {year} {2015})}\BibitemShut {NoStop}%
\bibitem [{\citenamefont {Kurumaji}\ \emph
  {et~al.}(2017{\natexlab{b}})\citenamefont {Kurumaji}, \citenamefont
  {Ishiwata},\ and\ \citenamefont {Tokura}}]{kurumaji_prb_2017}%
  \BibitemOpen
  \bibfield  {author} {\bibinfo {author} {\bibfnamefont {T.}~\bibnamefont
  {Kurumaji}}, \bibinfo {author} {\bibfnamefont {S.}~\bibnamefont {Ishiwata}},
  \ and\ \bibinfo {author} {\bibfnamefont {Y.}~\bibnamefont {Tokura}},\
  }\bibfield  {title} {\enquote {\bibinfo {title} {Diagonal magnetoelectric
  susceptibility and effect of {Fe} doping in the polar ferrimagnet
  {Mn$_2$Mo$_3$O$_8$}},}\ }\href {\doibase 10.1103/PhysRevB.95.045142}
  {\bibfield  {journal} {\bibinfo  {journal} {Phys. Rev. B}\ }\textbf {\bibinfo
  {volume} {95}},\ \bibinfo {pages} {045142} (\bibinfo {year}
  {2017}{\natexlab{b}})}\BibitemShut {NoStop}%
\bibitem [{\citenamefont {Kurumaji}\ \emph
  {et~al.}(2017{\natexlab{c}})\citenamefont {Kurumaji}, \citenamefont
  {Takahashi}, \citenamefont {Fujioka}, \citenamefont {Masuda}, \citenamefont
  {Shishikura}, \citenamefont {Ishiwata},\ and\ \citenamefont
  {Tokura}}]{kurumaji_prbr_2017}%
  \BibitemOpen
  \bibfield  {author} {\bibinfo {author} {\bibfnamefont {T.}~\bibnamefont
  {Kurumaji}}, \bibinfo {author} {\bibfnamefont {Y.}~\bibnamefont {Takahashi}},
  \bibinfo {author} {\bibfnamefont {J.}~\bibnamefont {Fujioka}}, \bibinfo
  {author} {\bibfnamefont {R.}~\bibnamefont {Masuda}}, \bibinfo {author}
  {\bibfnamefont {H.}~\bibnamefont {Shishikura}}, \bibinfo {author}
  {\bibfnamefont {S.}~\bibnamefont {Ishiwata}}, \ and\ \bibinfo {author}
  {\bibfnamefont {Y.}~\bibnamefont {Tokura}},\ }\bibfield  {title} {\enquote
  {\bibinfo {title} {Electromagnon resonance in a collinear spin state of the
  polar antiferromagnet {Fe$_2$Mo$_3$O$_8$}},}\ }\href {\doibase
  10.1103/PhysRevB.95.020405} {\bibfield  {journal} {\bibinfo  {journal} {Phys.
  Rev. B}\ }\textbf {\bibinfo {volume} {95}},\ \bibinfo {pages} {020405(R)}
  (\bibinfo {year} {2017}{\natexlab{c}})}\BibitemShut {NoStop}%
\bibitem [{\citenamefont {Watanabe}(1957)}]{watanabe_ptp_1957}%
  \BibitemOpen
  \bibfield  {author} {\bibinfo {author} {\bibfnamefont {Hiroshi}\ \bibnamefont
  {Watanabe}},\ }\bibfield  {title} {\enquote {\bibinfo {title} {{On the Ground
  Level Splitting of {Mn$^{++}$} and {Fe$^{+++}$} in Nearly Cubic Crystalline
  Field}},}\ }\href {\doibase 10.1143/PTP.18.405} {\bibfield  {journal}
  {\bibinfo  {journal} {Progress of Theoretical Physics}\ }\textbf {\bibinfo
  {volume} {18}},\ \bibinfo {pages} {405--420} (\bibinfo {year}
  {1957})}\BibitemShut {NoStop}%
\bibitem [{\citenamefont {McAlister}(1984)}]{mcalister_jap_1984}%
  \BibitemOpen
  \bibfield  {author} {\bibinfo {author} {\bibfnamefont {S.~P.}\ \bibnamefont
  {McAlister}},\ }\bibfield  {title} {\enquote {\bibinfo {title} {Unusual
  ferrimagnetism in {Mn$_2$Mo$_3$O$_8$} and {Sm$_2$In}},}\ }\href {\doibase
  10.1063/1.333657} {\bibfield  {journal} {\bibinfo  {journal} {Journal of
  Applied Physics}\ }\textbf {\bibinfo {volume} {55}},\ \bibinfo {pages}
  {2343--2345} (\bibinfo {year} {1984})}\BibitemShut {NoStop}%
\bibitem [{\citenamefont {Volkov}\ \emph {et~al.}(1985)\citenamefont {Volkov},
  \citenamefont {Goncharov}, \citenamefont {Kozlov}, \citenamefont {Lebedev},\
  and\ \citenamefont {Prokhorov}}]{volkov_infrared_1985}%
  \BibitemOpen
  \bibfield  {author} {\bibinfo {author} {\bibfnamefont {A.~A.}\ \bibnamefont
  {Volkov}}, \bibinfo {author} {\bibfnamefont {Yu.~G.}\ \bibnamefont
  {Goncharov}}, \bibinfo {author} {\bibfnamefont {G.~V.}\ \bibnamefont
  {Kozlov}}, \bibinfo {author} {\bibfnamefont {S.~P.}\ \bibnamefont {Lebedev}},
  \ and\ \bibinfo {author} {\bibfnamefont {A.~M.}\ \bibnamefont {Prokhorov}},\
  }\bibfield  {title} {\enquote {\bibinfo {title} {Dielectric measurements in
  the submillimeter wavelength region},}\ }\href {\doibase DOI:
  10.1016/0020-0891(85)90109-5} {\bibfield  {journal} {\bibinfo  {journal}
  {Infrared Phys.}\ }\textbf {\bibinfo {volume} {25}},\ \bibinfo {pages} {369}
  (\bibinfo {year} {1985})}\BibitemShut {NoStop}%
\bibitem [{\citenamefont {Fishman}\ \emph {et~al.}(2018)\citenamefont
  {Fishman}, \citenamefont {Fernandez-Baca},\ and\ \citenamefont
  {R{\~o}{\~o}m}}]{fishman_book_2018}%
  \BibitemOpen
  \bibfield  {author} {\bibinfo {author} {\bibfnamefont {Randy~S}\ \bibnamefont
  {Fishman}}, \bibinfo {author} {\bibfnamefont {Jaime~A}\ \bibnamefont
  {Fernandez-Baca}}, \ and\ \bibinfo {author} {\bibfnamefont {Toomas}\
  \bibnamefont {R{\~o}{\~o}m}},\ }\href {\doibase 10.1088/978-1-64327-114-9}
  {\emph {\bibinfo {title} {Spin-Wave Theory and its Applications to Neutron
  Scattering and THz Spectroscopy}}},\ 2053-2571\ (\bibinfo  {publisher}
  {Morgan {\&} Claypool Publishers},\ \bibinfo {year} {2018})\BibitemShut
  {NoStop}%
\bibitem [{\citenamefont {Kresse}\ and\ \citenamefont
  {Furthm\"uller}(1996{\natexlab{a}})}]{vasp1}%
  \BibitemOpen
  \bibfield  {author} {\bibinfo {author} {\bibfnamefont {G.}~\bibnamefont
  {Kresse}}\ and\ \bibinfo {author} {\bibfnamefont {J.}~\bibnamefont
  {Furthm\"uller}},\ }\bibfield  {title} {\enquote {\bibinfo {title}
  {Efficiency of \textit{ab-initio} total energy calculations for metals and
  semiconductors using a plane-wave basis set},}\ }\href {\doibase
  http://dx.doi.org/10.1016/0927-0256(96)00008-0} {\bibfield  {journal}
  {\bibinfo  {journal} {Computational Materials Science}\ }\textbf {\bibinfo
  {volume} {6}},\ \bibinfo {pages} {15} (\bibinfo {year}
  {1996}{\natexlab{a}})}\BibitemShut {NoStop}%
\bibitem [{\citenamefont {Kresse}\ and\ \citenamefont
  {Furthm\"uller}(1996{\natexlab{b}})}]{vasp2}%
  \BibitemOpen
  \bibfield  {author} {\bibinfo {author} {\bibfnamefont {G.}~\bibnamefont
  {Kresse}}\ and\ \bibinfo {author} {\bibfnamefont {J.}~\bibnamefont
  {Furthm\"uller}},\ }\bibfield  {title} {\enquote {\bibinfo {title} {Efficient
  iterative schemes for \textit{ab initio} total-energy calculations using a
  plane-wave basis set},}\ }\href {\doibase 10.1103/PhysRevB.54.11169}
  {\bibfield  {journal} {\bibinfo  {journal} {Phys. Rev. B}\ }\textbf {\bibinfo
  {volume} {54}},\ \bibinfo {pages} {11169} (\bibinfo {year}
  {1996}{\natexlab{b}})}\BibitemShut {NoStop}%
\bibitem [{\citenamefont {Duncan}\ \emph {et~al.}()\citenamefont {Duncan},
  \citenamefont {Tsurkan}, \citenamefont {Kezsmarki},\ and\ \citenamefont
  {Tsirlin}}]{duncan2020}%
  \BibitemOpen
  \bibfield  {author} {\bibinfo {author} {\bibfnamefont {D.}~\bibnamefont
  {Duncan}}, \bibinfo {author} {\bibfnamefont {V.}~\bibnamefont {Tsurkan}},
  \bibinfo {author} {\bibfnamefont {I.}~\bibnamefont {Kezsmarki}}, \ and\
  \bibinfo {author} {\bibfnamefont {A.A.}\ \bibnamefont {Tsirlin}},\
  }\href@noop {} {\ }\bibinfo {note} {(in preparation)}\BibitemShut {NoStop}%
\bibitem [{\citenamefont {Perdew}\ \emph {et~al.}(1996)\citenamefont {Perdew},
  \citenamefont {Burke},\ and\ \citenamefont {Ernzerhof}}]{perdew_prl_1996}%
  \BibitemOpen
  \bibfield  {author} {\bibinfo {author} {\bibfnamefont {John~P.}\ \bibnamefont
  {Perdew}}, \bibinfo {author} {\bibfnamefont {Kieron}\ \bibnamefont {Burke}},
  \ and\ \bibinfo {author} {\bibfnamefont {Matthias}\ \bibnamefont
  {Ernzerhof}},\ }\bibfield  {title} {\enquote {\bibinfo {title} {Generalized
  gradient approximation made simple},}\ }\href {\doibase
  10.1103/PhysRevLett.77.3865} {\bibfield  {journal} {\bibinfo  {journal}
  {Phys. Rev. Lett.}\ }\textbf {\bibinfo {volume} {77}},\ \bibinfo {pages}
  {3865--3868} (\bibinfo {year} {1996})}\BibitemShut {NoStop}%
\bibitem [{\citenamefont {Nath}\ \emph {et~al.}(2014)\citenamefont {Nath},
  \citenamefont {Ranjith}, \citenamefont {Roy}, \citenamefont {Johnston},
  \citenamefont {Furukawa},\ and\ \citenamefont {Tsirlin}}]{nath2014}%
  \BibitemOpen
  \bibfield  {author} {\bibinfo {author} {\bibfnamefont {R.}~\bibnamefont
  {Nath}}, \bibinfo {author} {\bibfnamefont {K.~M.}\ \bibnamefont {Ranjith}},
  \bibinfo {author} {\bibfnamefont {B.}~\bibnamefont {Roy}}, \bibinfo {author}
  {\bibfnamefont {D.~C.}\ \bibnamefont {Johnston}}, \bibinfo {author}
  {\bibfnamefont {Y.}~\bibnamefont {Furukawa}}, \ and\ \bibinfo {author}
  {\bibfnamefont {A.~A.}\ \bibnamefont {Tsirlin}},\ }\bibfield  {title}
  {\enquote {\bibinfo {title} {Magnetic transitions in the spin-5/2 frustrated
  magnet {BiMn$_2$PO$_6$} and strong lattice softening in {BiMn$_2$PO$_6$} and
  {BiZn$_2$PO$_6$} below {200 K}},}\ }\href {\doibase
  10.1103/PhysRevB.90.024431} {\bibfield  {journal} {\bibinfo  {journal} {Phys.
  Rev. B}\ }\textbf {\bibinfo {volume} {90}},\ \bibinfo {pages} {024431}
  (\bibinfo {year} {2014})}\BibitemShut {NoStop}%
\bibitem [{\citenamefont {Xiang}\ \emph {et~al.}(2011)\citenamefont {Xiang},
  \citenamefont {Kan}, \citenamefont {Wei}, \citenamefont {Whangbo},\ and\
  \citenamefont {Gong}}]{xiang2011}%
  \BibitemOpen
  \bibfield  {author} {\bibinfo {author} {\bibfnamefont {H.~J.}\ \bibnamefont
  {Xiang}}, \bibinfo {author} {\bibfnamefont {E.~J.}\ \bibnamefont {Kan}},
  \bibinfo {author} {\bibfnamefont {S.-H.}\ \bibnamefont {Wei}}, \bibinfo
  {author} {\bibfnamefont {M.-H.}\ \bibnamefont {Whangbo}}, \ and\ \bibinfo
  {author} {\bibfnamefont {X.~G.}\ \bibnamefont {Gong}},\ }\bibfield  {title}
  {\enquote {\bibinfo {title} {Predicting the spin-lattice order of frustrated
  systems from first principles},}\ }\href {\doibase
  10.1103/PhysRevB.84.224429} {\bibfield  {journal} {\bibinfo  {journal} {Phys.
  Rev. B}\ }\textbf {\bibinfo {volume} {84}},\ \bibinfo {pages} {224429}
  (\bibinfo {year} {2011})}\BibitemShut {NoStop}%
\bibitem [{\citenamefont {Turov}(1965)}]{turov_book_1965}%
  \BibitemOpen
  \bibfield  {author} {\bibinfo {author} {\bibfnamefont {E.~A.}\ \bibnamefont
  {Turov}},\ }\href@noop {} {\emph {\bibinfo {title} {Physical Properties of
  Magnetically Ordered Crystals. Translated From the Russian by Scripta
  Technica, Inc. Translation Edited by A. Tybulewicz and S. Chomet}}}\
  (\bibinfo  {publisher} {Academic Press},\ \bibinfo {year} {1965})\BibitemShut
  {NoStop}%
\bibitem [{\citenamefont {Odom}\ \emph {et~al.}(2006)\citenamefont {Odom},
  \citenamefont {Hanneke}, \citenamefont {D'Urso},\ and\ \citenamefont
  {Gabrielse}}]{odom_prl_2007}%
  \BibitemOpen
  \bibfield  {author} {\bibinfo {author} {\bibfnamefont {B.}~\bibnamefont
  {Odom}}, \bibinfo {author} {\bibfnamefont {D.}~\bibnamefont {Hanneke}},
  \bibinfo {author} {\bibfnamefont {B.}~\bibnamefont {D'Urso}}, \ and\ \bibinfo
  {author} {\bibfnamefont {G.}~\bibnamefont {Gabrielse}},\ }\bibfield  {title}
  {\enquote {\bibinfo {title} {New measurement of the electron magnetic moment
  using a one-electron quantum cyclotron},}\ }\href {\doibase
  10.1103/PhysRevLett.97.030801} {\bibfield  {journal} {\bibinfo  {journal}
  {Phys. Rev. Lett.}\ }\textbf {\bibinfo {volume} {97}},\ \bibinfo {pages}
  {030801} (\bibinfo {year} {2006})}\BibitemShut {NoStop}%
\bibitem [{tor()}]{torque}%
  \BibitemOpen
  \href@noop {} {}\bibinfo {note} {Using the
  $\theta_M=\frac{\theta_0}{H-H_{C1}}$ relation corresponding to
  Fig.~\ref{fig_magn}(c), in a simplified model $M=H\cos \theta_M$ is the
  magnitude of the magnetization and $\tau = H^2 \sin 2 \theta_M$ is that of
  the torque for $H>H_{C1}$ magnetic fields along $c$. In the $\theta_M \ll 1$
  limit the local miminum of the torque at high fields is $\tau_{min}
  =8\theta_0 H_{C1}$ at $H=2H_{C1}$, while its peak value is $\tau_{max}
  =H_{C1}^2$. Thus, using the observed $\frac{\tau_{min}}{\tau_{max}}\approx
  0.5$ relation (Fig.~\ref{fig_torque}(a)), the full width at half maximum of
  the peak in the field dependence of torque can be estimated as $\Delta \tau =
  (2-\sqrt{3})H_{C1}$ }\BibitemShut {NoStop}%
\bibitem [{\citenamefont {Szaller}\ \emph {et~al.}(2017)\citenamefont
  {Szaller}, \citenamefont {Kocsis}, \citenamefont {Bord\'acs}, \citenamefont
  {Feh\'er}, \citenamefont {R{\~o}{\~o}m}, \citenamefont {Nagel}, \citenamefont
  {Engelkamp}, \citenamefont {Ohgushi},\ and\ \citenamefont
  {K\'ezsm\'arki}}]{szaller_prb_2017}%
  \BibitemOpen
  \bibfield  {author} {\bibinfo {author} {\bibfnamefont {D.}~\bibnamefont
  {Szaller}}, \bibinfo {author} {\bibfnamefont {V.}~\bibnamefont {Kocsis}},
  \bibinfo {author} {\bibfnamefont {S.}~\bibnamefont {Bord\'acs}}, \bibinfo
  {author} {\bibfnamefont {T.}~\bibnamefont {Feh\'er}}, \bibinfo {author}
  {\bibfnamefont {T.}~\bibnamefont {R{\~o}{\~o}m}}, \bibinfo {author}
  {\bibfnamefont {U.}~\bibnamefont {Nagel}}, \bibinfo {author} {\bibfnamefont
  {H.}~\bibnamefont {Engelkamp}}, \bibinfo {author} {\bibfnamefont
  {K.}~\bibnamefont {Ohgushi}}, \ and\ \bibinfo {author} {\bibfnamefont
  {I.}~\bibnamefont {K\'ezsm\'arki}},\ }\bibfield  {title} {\enquote {\bibinfo
  {title} {Magnetic resonances of multiferroic {TbFe}$_3${(BO}$_3$)$_4$},}\
  }\href {\doibase 10.1103/PhysRevB.95.024427} {\bibfield  {journal} {\bibinfo
  {journal} {Phys. Rev. B}\ }\textbf {\bibinfo {volume} {95}},\ \bibinfo
  {pages} {024427} (\bibinfo {year} {2017})}\BibitemShut {NoStop}%
\bibitem [{\citenamefont {Peedu}\ \emph {et~al.}(2019)\citenamefont {Peedu},
  \citenamefont {Kocsis}, \citenamefont {Szaller}, \citenamefont {Viirok},
  \citenamefont {Nagel}, \citenamefont {R{\~o}{\~o}m}, \citenamefont {Farkas},
  \citenamefont {Bord\'acs}, \citenamefont {Kamenskyi}, \citenamefont
  {Zeitler}, \citenamefont {Tokunaga}, \citenamefont {Taguchi}, \citenamefont
  {Tokura},\ and\ \citenamefont {K\'ezsm\'arki}}]{peedu_prb_2019}%
  \BibitemOpen
  \bibfield  {author} {\bibinfo {author} {\bibfnamefont {L.}~\bibnamefont
  {Peedu}}, \bibinfo {author} {\bibfnamefont {V.}~\bibnamefont {Kocsis}},
  \bibinfo {author} {\bibfnamefont {D.}~\bibnamefont {Szaller}}, \bibinfo
  {author} {\bibfnamefont {J.}~\bibnamefont {Viirok}}, \bibinfo {author}
  {\bibfnamefont {U.}~\bibnamefont {Nagel}}, \bibinfo {author} {\bibfnamefont
  {T.}~\bibnamefont {R{\~o}{\~o}m}}, \bibinfo {author} {\bibfnamefont {D.~G.}\
  \bibnamefont {Farkas}}, \bibinfo {author} {\bibfnamefont {S.}~\bibnamefont
  {Bord\'acs}}, \bibinfo {author} {\bibfnamefont {D.~L.}\ \bibnamefont
  {Kamenskyi}}, \bibinfo {author} {\bibfnamefont {U.}~\bibnamefont {Zeitler}},
  \bibinfo {author} {\bibfnamefont {Y.}~\bibnamefont {Tokunaga}}, \bibinfo
  {author} {\bibfnamefont {Y.}~\bibnamefont {Taguchi}}, \bibinfo {author}
  {\bibfnamefont {Y.}~\bibnamefont {Tokura}}, \ and\ \bibinfo {author}
  {\bibfnamefont {I.}~\bibnamefont {K\'ezsm\'arki}},\ }\bibfield  {title}
  {\enquote {\bibinfo {title} {Spin excitations of magnetoelectric
  $\mathrm{LiNiPO}{}_{4}$ in multiple magnetic phases},}\ }\href {\doibase
  10.1103/PhysRevB.100.024406} {\bibfield  {journal} {\bibinfo  {journal}
  {Phys. Rev. B}\ }\textbf {\bibinfo {volume} {100}},\ \bibinfo {pages}
  {024406} (\bibinfo {year} {2019})}\BibitemShut {NoStop}%
\end{thebibliography}
%

\end{document}